\documentclass{article}

\usepackage{arxiv}

\usepackage{natbib}
\usepackage{graphicx}
\usepackage{color}
\usepackage{amsmath, amssymb}
\usepackage[utf8]{inputenc} 
\usepackage[T1]{fontenc}    
\usepackage{hyperref}       
\usepackage{url}            
\usepackage{booktabs}       
\usepackage{amsfonts}       
\usepackage{nicefrac}       
\usepackage{microtype}      
\usepackage{lipsum}

\usepackage{txfonts}

\graphicspath{{./}}

\title{Review of the particle background of the Athena X-IFU instrument}

\author{
  David S.~Hippocampus\thanks{Use footnote for providing further
    information about author (webpage, alternative
    address)---\emph{not} for acknowledging funding agencies.} \\
  Department of Computer Science\\
  Cranberry-Lemon University\\
  Pittsburgh, PA 15213 \\
  \texttt{hippo@cs.cranberry-lemon.edu} \\
   \And}

\author{Simone Lotti \\
Istituto di Astrofisica e Planetologia Spaziali,\\
Via fosso del cavaliere 100,\\
Roma, Italia\\
\texttt{simone.lotti@inaf.it}\\

\And

Matteo D'Andrea\\

Istituto di Astrofisica e Planetologia Spaziali,\\
Via fosso del cavaliere 100,\\
Roma, Italia\\

\And

Silvano Molendi \\
Istituto di Astrofisica Spaziale e Fisica Cosmica di Milano,\\
Via Alfonso Corti 12,\\
20133 Milano, Italia \\

\And

Claudio Macculi\\
Istituto di Astrofisica e Planetologia Spaziali,\\
Via fosso del cavaliere 100,\\
Roma, Italia\\

\And
Gabriele Minervini\\
Istituto di Astrofisica e Planetologia Spaziali,\\
Via fosso del cavaliere 100,\\
Roma, Italia\\

\And

Valentina Fioretti\\
Osservatorio di Astrofisica e Scienza dello Spazio di Bologna, \\
Via Piero Gobetti, 93/3,\\
Bologna, Italia\\

\And

Monica Laurenza\\
Istituto di Astrofisica e Planetologia Spaziali,\\
Via fosso del cavaliere 100,\\
Roma, Italia\\

\And

Christian Jacquey\\
IRAP, Université de Toulouse, \\
CNRS, CNES, UPS,\\
Toulouse, France\\

\And
Luigi Piro\\
Istituto di Astrofisica e Planetologia Spaziali,\\
Via fosso del cavaliere 100,\\
Roma, Italia\\
}

\begin{document}
\maketitle
\begin{abstract}

X-ray observations are limited by the background, due to intrinsic faintness or diffuse nature of the sources. The future Athena X-ray observatory has among its goals the characterization of these sources.
We aim at characterizing the particle-induced background of the Athena microcalorimeter, in both its low (Soft Protons) and high (GCR) energy induced components, to assess the instrument capability to characterize background dominated sources such as the outskirts of clusters of galaxies. We compare two radiation environments, namely the L1 and L2 Lagrangian points, and derive indications against the latter. We estimate the particle-induced background level on the X-IFU microcalorimeter with Monte Carlo simulations, before and after all the solutions adopted to reduce its level. Concerning the GCR induced component the background level is compliant with the mission requirement. Regarding the Soft Protons component, the analysis does not predict dramatically different backgrounds in the L1 and L2 orbits. However, the lack of data concerning the L2 environment labels it as very weakly characterizable, and thus we advise against its choice as orbit for X-ray missions.
We then use these background levels to simulate the observation of a typical galaxy cluster from its center out to 1.2 $R_{200}$ to probe the characterization capabilities of the instrument out to the outskirts. We find that without any background reduction it is not possible to characterize the properties of the cluster in the outer regions. We also find no improvement of the observations when carried out during the solar maximum with respect to the solar minimum conditions.

\end{abstract}


\section{Introduction} \label{sec:intro}\label{sec:1}

 Athena \citep{whitepaper} is the second large-class X-ray mission of the European Space Agency Cosmic Vision program, with a launch foreseen on early 2030 towards an L2 halo orbit and dedicated to the study of the hot and energetic universe. 
 The mission couples a high-performance X-ray Telescope (1.4 $m^2$ at 1 keV, and an angular resolution of 5 arcsec) with two complementary focal-plane instruments, the Wide Field Imager (WFI) optimized for surveys \citep{wfi}, and the X-ray Integral Field Unit (X-IFU), providing integral field spatially resolved high energy resolution spectroscopy (2.5 eV at 6 keV) over a 5 arcmin FoV in the 0.2-12 keV energy band, thanks to an array of $\sim 3840$ transition edge sensors (TES) microcalorimeters operated at 50 mK with 249 $\mu m$ pitch for a 2.3 $cm^2$ sensitive area \citep{barret2018}.

 X-ray observations are usually severely limited by the background, due to the intrinsic faintness of the astrophysical sources involved or to their diffuse nature, and this is especially true for Athena, which has among its goals the characterization of faint/distant/diffuse sources unobservable with any other present or planned instrument \citep{whitepaper}. For this reason substantial effort has been dedicated to the assessment of the expected background level.
 
 In this paper we address the particle-induced background of the X-IFU instrument, and assess the capability of the mission to characterize background-dominated sources, discussing the specific case of the outskirts of clusters of galaxies. 
 
This paper is structured as follows: in Section \ref{sec:2} we introduce general issues on the background of X-ray instruments, in Section \ref{sec:3} we focus on the methodology used to estimate the particle background induced on the X-IFU. In Section \ref{sec:4} we report the results of the high-energy induced component of the particle background, while in Section \ref{sec:5} we do the same for the low energy induced component. In Section \ref{sec:6} we use the background levels calculated in the previous sections to estimate the X-IFU observation capabilities of a galaxy cluster outskirt out to $1.2R_{200}$, and in the final section are reported our conclusions.


\section{The background of X-ray instruments}\label{sec:2}

X-ray instruments are subject to different kinds of background, of different nature and features. 

The first component, dominant at $E<2-3$ keV, is induced by the Cosmic X-ray Background (CXB), a diffuse X-ray emission observed in every direction. 
At low energies ($<1$ keV) it is composed of several contributions, each with its own spatial and temporal variations, while at higher energies it is generated by the unresolved emission from AGNs. This component is m induced by photons produced by X-ray sources in the sky, it reaches the detector from the optics and can be damped only resolving the AGN-induced component into the single sources that create it.

The second component is generated by charged particles, and can be further divided in two separate contributions:
\begin{itemize}
 \item Low energy particles (Soft Protons - SP), mostly protons below 200 keV, that follow a path similar to the CXB, being concentrated by the mirrors into the focal plane, and releasing all of their remaining energy in the detectors. 
 \item High energy particles (Galactic Cosmic Rays - GCR), that possess enough energy to travel through the spacecraft and reach the detector from any direction, tipically releasing inside it a small fraction of their energy. These particles also create showers of secondary particles (mostly electrons) along the way, that can also induce additional background.
\end{itemize}
In this paper we deal with both the contributions of the second component, while for the first we assume the modelization reported in \citet{lotti2014}.

\section{The particle background estimation}\label{sec:3}
There are no data available for X-Ray microcalorimeters in L2 (or L1) at the moment of writing, thus it is impossible to estimate the expected background level relying on existing data. The problem is also too complex to face using an analytical approach, and consequently the background estimates are performed using Monte Carlo simulations with the Geant4 software \citep{geant1,geant2,geant3}.

Geant4 is a toolkit that provides access to several different models to reproduce physical processes the user is interested in. Also, it is possible to set the simulation detail level to suit the user needs, balancing the accuracy of the results with the simulation speed.

Thanks to this tool it is possible to create a virtual model of the instrument and its surroundings (the so-called Mass Model) and of the particle environment it will be placed into, and analyze how they interact with each other and ultimately, given the proper settings, to predict the expected instrument performances, among which the background level. 
Furthermore, it is possible to test the effect of changes to the instrument mass model (sensitivity analysis) and identify the configuration that optimizes the detector performances.

We have exploited this Monte Carlo tool to estimate the impact of both the GCR-induced component and the SP-induced one.

\subsection{Input spectra}\label{sec:spectra}

 \begin{figure*}
 \centering
 \includegraphics[width=0.49\textwidth]{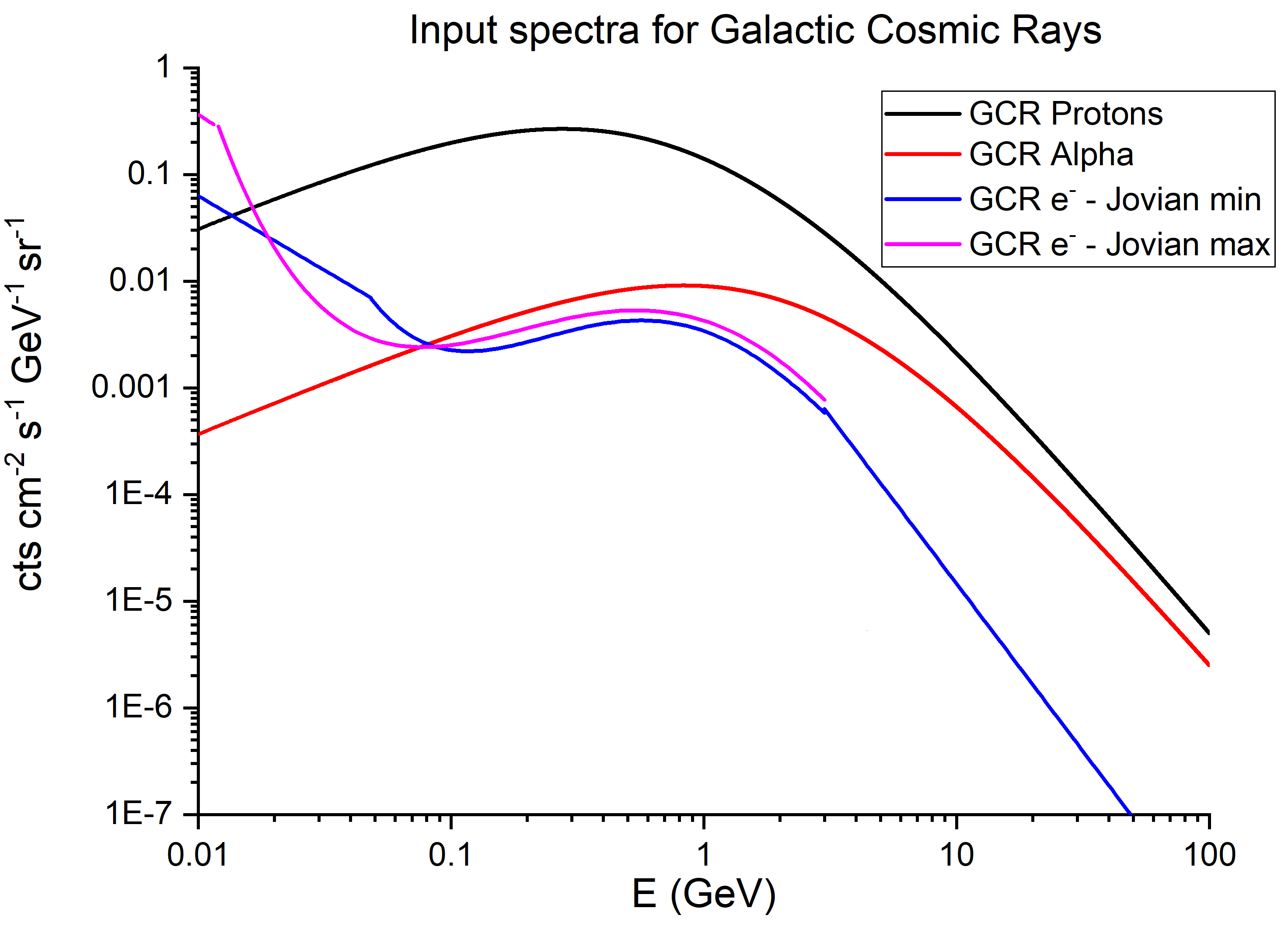}
 \includegraphics[width=0.47\textwidth]{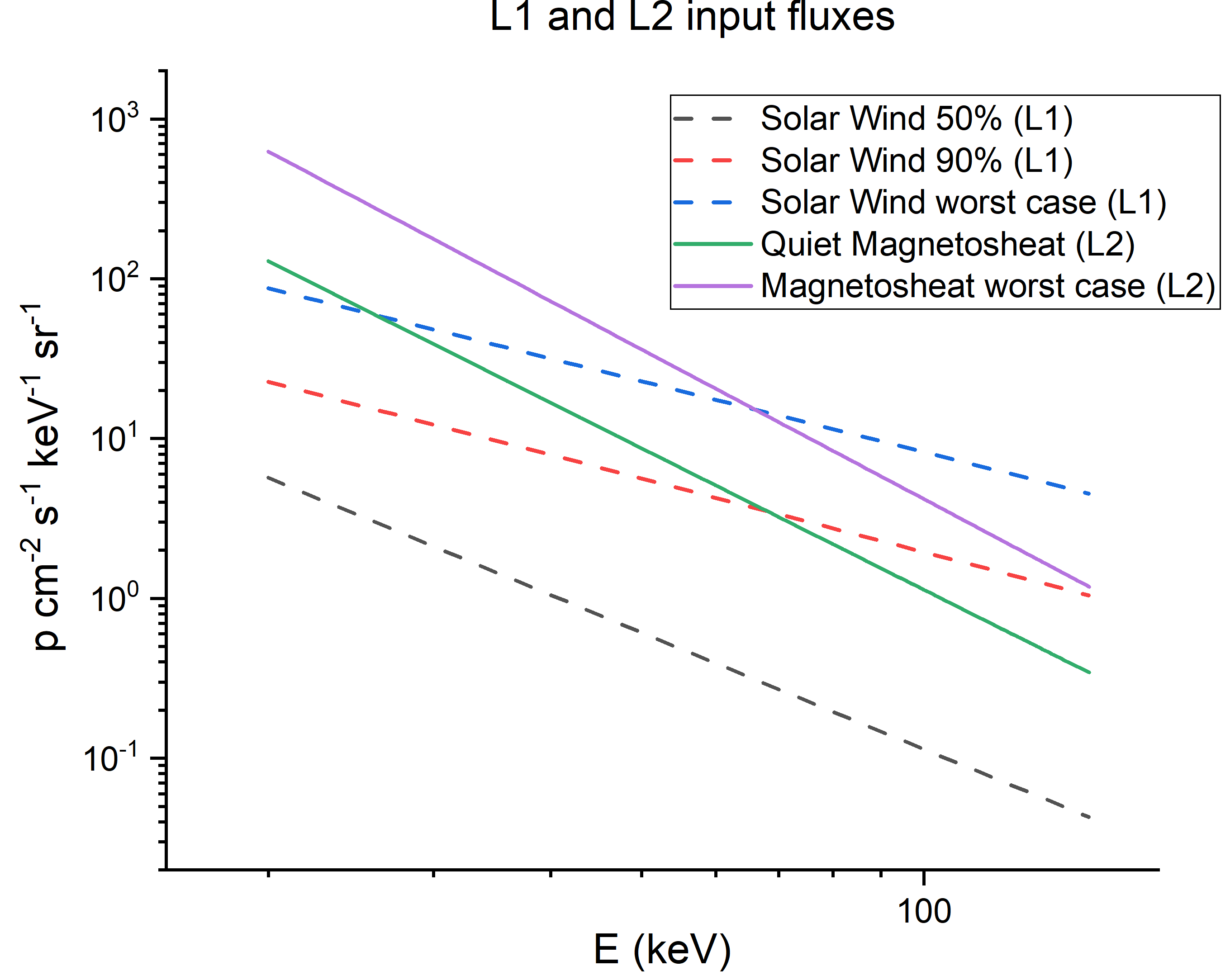}

 \caption{Left: input spectra for GCR protons, GCR alpha particles and electrons (both GCR and Jovian contribution). Right: L1 and L2 input fluxes for solar wind (the most probable environment), and for the worst case foreseen in both environments. See text for details.}
 \label{fig:inputspectra}
 \end{figure*}

The Athena mission is foreseen to fly in the L2 Lagrangian point. However, it is under discussion the possibility to direct the spacecraft in the L1 point. We report here the results of the analysis of the environment of the 2 orbits.

Regarding the GCR component, it changes through the solar system with the distance from the Sun, and L1 and L2 are close enough to not expect any significant difference between the 2 orbits (${\Delta R}/{<R>}\sim2\%$, with R the distance from the Sun). The reference spectra are listed in the following, and shown in Figure \ref{fig:inputspectra}.

The reference model for GCR protons has been described in \citet{usoskin2005}, here we report the formula for the differential intensity J of cosmic ray nucleon at 1 AU in particles $cm^{-2} s^{-1} sr^{-1} GeV^{-1}$:

\begin{equation}
J(E,\phi)=J_{LIS}(E+\phi)\frac{E(E+2E_r)}{(E+\phi)(E+\phi+2E_r)} 
\end{equation}

where
\begin{equation}
J_{LIS}(E)=\frac{1.9\cdot[E(E+2E_r)]^{-1.39}}{1+0.4866\cdot[E\cdot(E+2E_r)]^{-1.255}}
\end{equation}

is the Local Interstellar Spectrum (LIS) of cosmic ray nuclei. E is the kinetic energy of the particle in GeV, $E_r$ = 0.938 GeV is the proton rest mass, and $\phi$ is the modulation potential in GV, also known as the "force field parameter" and the "modulation strength", and represents the mean energy loss of the GCR particle inside the heliosphere. 
We used this model to fit satellites data for 2009 from SOHO \citep{SOHO}, Pamela \citep{pamela} and Voyager, obtaining  $\phi$ = 0.3793 $\pm$ 0.0014 GV for the GCR maximum (solar minimum). 
We chose 2009 as reference year since it corresponds to a solar minumum in a solar cycle with the same negative polarity as the one in which Athena will operate \citep{pamela} and we use it as a proxy for year 2031 in which the maximum GCR flux is expected for cycle 26. This spectrum was checked against the Heliospheric Modulation Model \citep{helmod} results for 2009, and resulted in agreement for $E>1$ GeV and conservative at energies below.
For what concerns the GCR minimum (solar maximum) to be possibly experienced by Athena during its lifetime, we kept using solar cycle 24 as a proxy for the solar cycle 26 in which Athena will fly, and used the year 2014 as solar maximum to be experienced by Athena during its lifetime, fitting SOHO data to obtain an expected $\phi$ = 0.803 $\pm$ 0.01 GV value as GCR minimum in that timeframe.

Alpha particles spectrum has been evaluated from Kuznetzov empirical model \citep{alpha} for quiet Sun conditions: 
 \begin{equation}
F_{He}=1.085\times10^5\cdot E^{-2.72} \left( \frac{E}{E+2304} \right) ^{3.7}
\end{equation}
where E is the kinetic energy per nucleus in MeV and $F_{He}$ is in particles $cm^{-2} s^{-1} sr^{-1} MeV^{-1}$. The adopted spectra correspond to quiet sun conditions, where the Sun magnetic shielding effectiveness is at its lowest, thus providing the maximum Alpha particles flux.
The adopted Alpha particles spectrum was checked against the Heliospheric Modulation Model \citep{helmod,helmodalpha,helmodalpha2} results for 2009, and resulted in perfect agreement.


The high energy end ($>3$ GeV) of the GCR electrons spectrum was fitted as a powerlaw from Pamela data for 2009 \citep{pamela-ele} as:
\begin{equation}
F_{e^-}=200\cdot E^{-3.14}
\end{equation}
Where E is the kinetic energy in GeV and $F_{e^-}$ is in particles $m^{-2} s^{-1} sr^{-1} GeV^{-1}$.
\
At lower energies we decided to take into account the role of interplanetary (Jovian) electrons: we fitted the spectra reported in \citet{Grimani2009} for a negative polarity period, for the maximum Jovian component:
\begin{equation}
F_{e^-}^{J_{max}}=\begin{cases}
3.56\cdot E^{-1.5}, \hspace{80pt minus 1fil} \text{if $E< 0.0115$ GeV} \hfilneg \\
exp(-3.75-0.76\cdot log(E)-\\-0.69\cdot log^2(E) + 0.057\cdot log^3(E)+\\+0.025\cdot log^4(E)), \hspace{40pt minus 1fil} \text{if $0.0115<E<3$ GeV} \hfilneg \\
\end{cases}
\end{equation}

and for the minimum Jovian component:

\begin{equation}
F_{e^-}^{J_{min}}=\begin{cases}
1.02\cdot E^{-1.4}, \hspace{80pt minus 1fil} \text{if $E< 0.0483$ GeV} \hfilneg \\
1500\cdot exp(-3.78-0.83\cdot log(E)-\\-0.77\cdot log^2(E) + 0.0045\cdot log^3(E)+\\+0.061\cdot log^4(E)), \hspace{40pt minus 1fil} \text{if $0.0483<E<3$ GeV} \hfilneg \\
\end{cases}
\end{equation}

This spectra were checked against the Heliospheric Modulation Model \citep{helmod,helmodele} results for 2009, and our modelization choice resulted in agreement.

According to the latest formulation, the Athena background requirement should be satisfied with the assumption of an external reference flux equal to 80\% of the GCR maximum flux, which is what we refer to in all of the following.


Once established the largest flux conditions, we investigated the GCR protons flux expected variability from cycle to cycle. There is a correlation among the Neutron Monitors count rate on Earth \citep{oulu}, that are part of long-term cosmic ray measurements taken from the Earth's surface, and the integrated GCR protons flux, so we analyzed the Neutron Monitor count rate collected from 1964 up to 2017 to establish variability from cycle to cycle. During this time span, 5 different solar cycles occurred: three of them characterized by a heliosphere with a negative polarity (solar minima in 1965, 1987 and 2009), such as the one expected in the Athena lifetime \citep{pamela}. Analyzing the Neutron Monitor counts over a 2 years average around the solar minima, the GCR flux indetermination is $\sim 26\%$. This estimate is however based on a very small sample (3 solar minima with negative polarity since 1964). If we compute the population variance at 95\% confidence level the resulting indetermination can be up to six times larger than the sample variance.

Regarding the Soft Protons component, which belongs in the 100s of keV regime, important efforts have been spent in the characterization of the two possibilities under discussion for the Athena orbit, namely the L1 and L2 lagrangian points of the Earth-Sun system. The spectral shapes for the Soft Protons (SP) detected by several satellites present in L1 and L2 \citep{laurenza2019,lotti2018} are reported in Table \ref{tab:SPfluxes}. The integrated fluxes in the 40-80 keV are reported in Table \ref{tab:SPresume}.

Observations of the ARTEMIS probes \citep{artemis} simultaneously obtained in both solar wind and distant magnetotail have shown that the L2 soft proton fluxes are similar to the ones measured at L1 but with the superimposition of an important component fed by the particles accelerated inside the magnetosphere as soon as there is some geomagnetic activity, i.e. very often.

\begin{table}[]
 \caption{Proton flux fit functions for different thresholds, valid for the 50 - 5000 keV energy range. $E$ is the kinetic energy in keV, flux is in particles $cm^{-2} s^{-1} sr^{-1} keV^{-1}$. SW stands for Solar Wind, while MS for Magnetosheath.}
 \label{tab:SPfluxes}
\begin{tabular}{ll}
 \hline
 Cumulative probability & Fit function \\
 \noalign{\smallskip}
 \hline
 \noalign{\smallskip}
 Solar Wind 50\% (L1) & $8444\cdot E^{-2.44} e^{-2.4\times 10^{-4}E}$ \\
 Solar Wind 90\% (L1) & $2096\cdot E^{-1.51} e^{-2.6\times 10^{-4}E}$ \\
 Solar Wind worst case (L1) & $6734\cdot E^{-1.45} e^{-2.67\times 10^{-4}E} $ \\
 \noalign{\smallskip}
 Quiet Magnetosheath (L2) & $8.6\times10^5\cdot E^{-2.94}$ \\
 Magnetosheath worst case (L2) & $7\times 10^6 \cdot E^{-3.11}$ 

\end{tabular}
\end{table}

We emphasize that all the available data related to the L2 environment comprises only a few months of data $close$ to the L2 point in a 2 years period, i.e. only covering a small fraction of a single solar cycle. Moreover we point out that these 2 years occurred during the solar cycle decreasing phase. Due to this very limited coverage the knowledge of the soft proton fluxes in L2 is much more limited with respect to the one inferred from the corresponding L1 dataset, which involves tens of years of data along two solar cycles. Consequently, despite having performed the work for both cases, it is not possible to produce a reliable forecast of the L2 SP fluxes, and the L1 estimates should be considered the only reliable ones.

\subsection{X-IFU mass model}\label{sec:massmodel}
 \begin{figure*}
 \centering
 \includegraphics[width=0.48\textwidth]{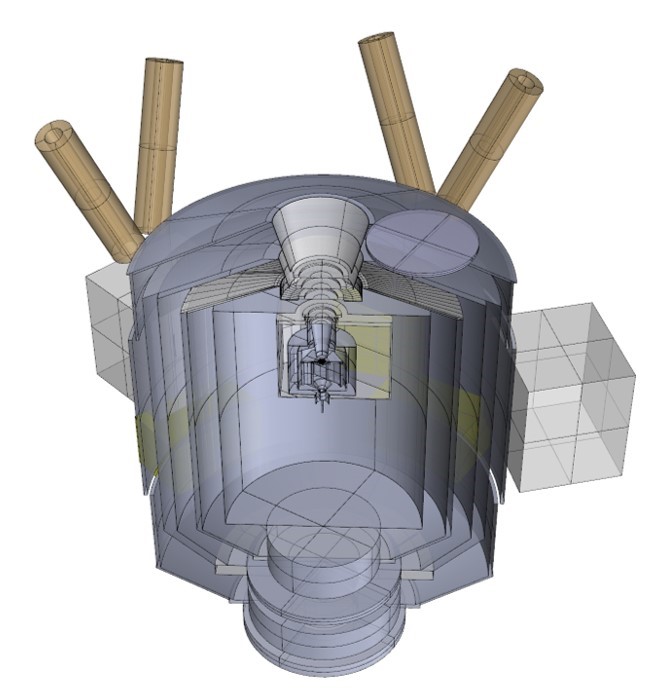}
 \includegraphics[width=0.48\textwidth]{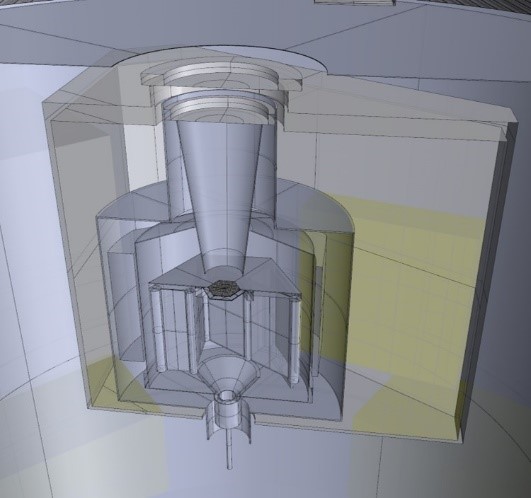}

 \caption{CAD of the simplified mass models used in the Geant4 simulations: cryostat (left) and FPA (right).}
 \label{fig:massmodels}
 \end{figure*}
In 2017 we received updated drawings of both the cryostat from CNES (Centre National D'Etudes Spatiales) and the FPA (Focal Plane Assembly) from SRON (Netherlands Institute for Space Research), respectively, from which we derived an updated version of the Geant4 mass model of the X-IFU. The engineering CAD models received were far too detailed and complex for the insertion in the Monte Carlo code, so a substantial work of simplification has been carried out, stripping off the model of unnecessary details while preserving the masses and shapes as much as possible (see Figure \ref{fig:massmodels}). The level of detail increases closer to the detectors, which are modeled in the greatest detail (all the $\sim 3840$ pixels of the main detector are present and modeled according to the latest layout available at that time).

In the basic FPA configuration the detector is directly exposed to the Niobium cryogenic shielding, but in the latest baseline a passive shielding was introduced in order to reduce the flux of secondary particles. We investigated the background expected on the detector in both configurations.

\section{The background foreseen for the X-IFU: the high energy component}\label{sec:4}

As the GCR protons are the dominant component of the external particle flux when compared to electrons and alpha particles (and of the residual background, as we will see in the following), we perform most of the analysis on this component, and once the optimal configuration for the background has been established we simulate also the other particle species.

\begin{figure*}
 \centering
 \includegraphics[width=0.7\textwidth]{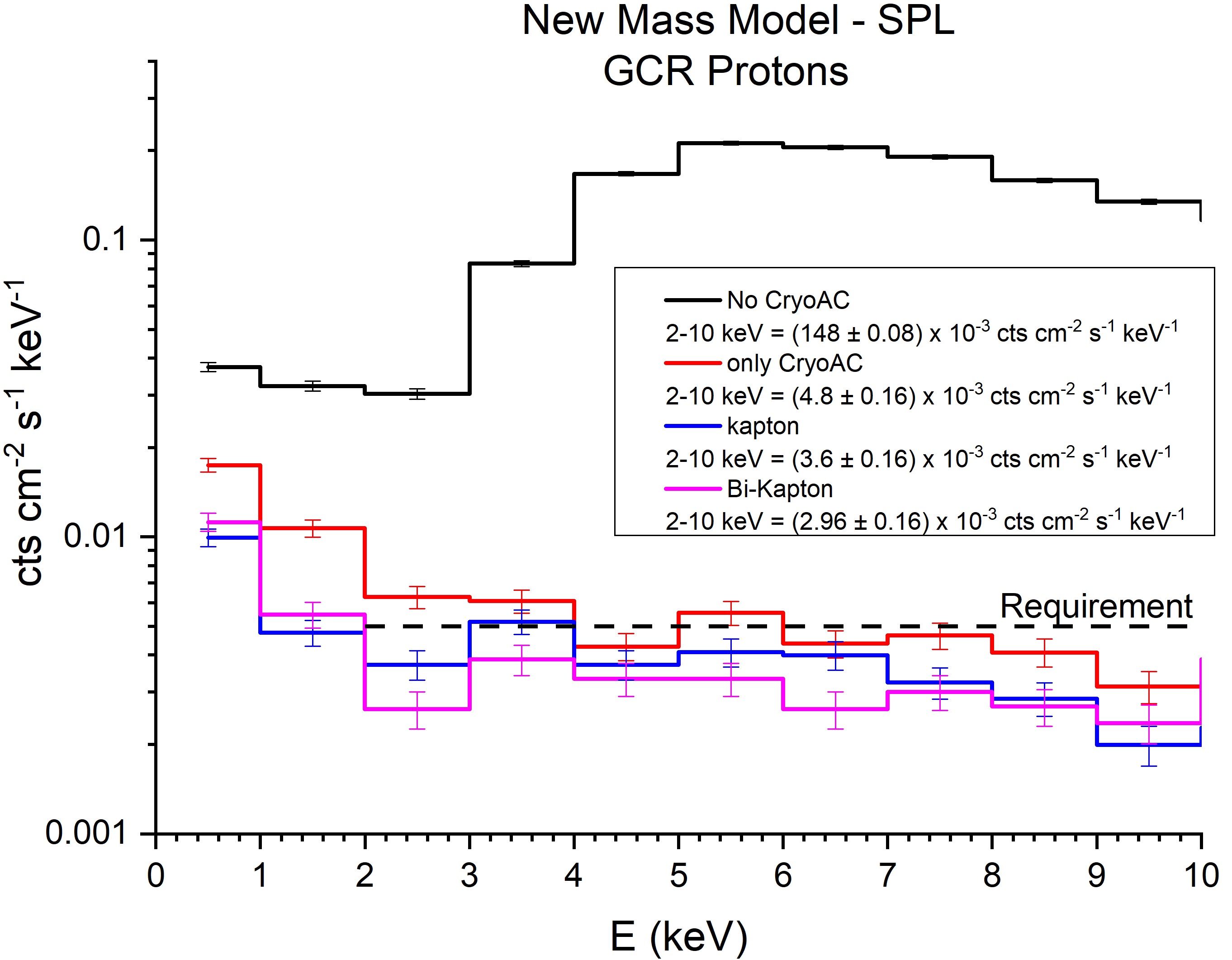}
 \caption{Background levels in different FPA configurations.}
 \label{fig:bkglevels}
 \end{figure*}

Without any kind of reduction technique the X-IFU would experience a GCR-protons induced background level of $(148 \pm 0.08) \times 10^{-3}$ cts $cm^{-2}s^{-1}keV^{-1}$ in the 2-10 keV energy band, 30 times above the $5 \times 10^{-3}$ cts $cm^{-2}s^{-1}keV^{-1}$ requirement (see Figure \ref{fig:bkglevels}). For this reason, a cryogenic anticoincidence detector (CryoAC) was introduced in the baseline configuration. This reduced the unrejected background level down to $(4.8 \pm 0.16) \times 10^{-3}$ cts $cm^{-2}s^{-1}keV^{-1}$, efficiently discriminating against high energy particles that can cross the main detector and reach the CryoAC. Assuming a 20\% of margin due to the systematic uncertainty on Geant4 reproducibility of in-flight background data, as found by the accuracy of the simulations \citep{fioretti2018}, the background level is 20\% above the requirement. This evaluation was performed taking into account only GCR protons (which produce $\sim70\%$ of the residual particle background), while later on we report the expected full contribution.

\begin{figure*}
 \centering
 \includegraphics[width=0.48\textwidth]{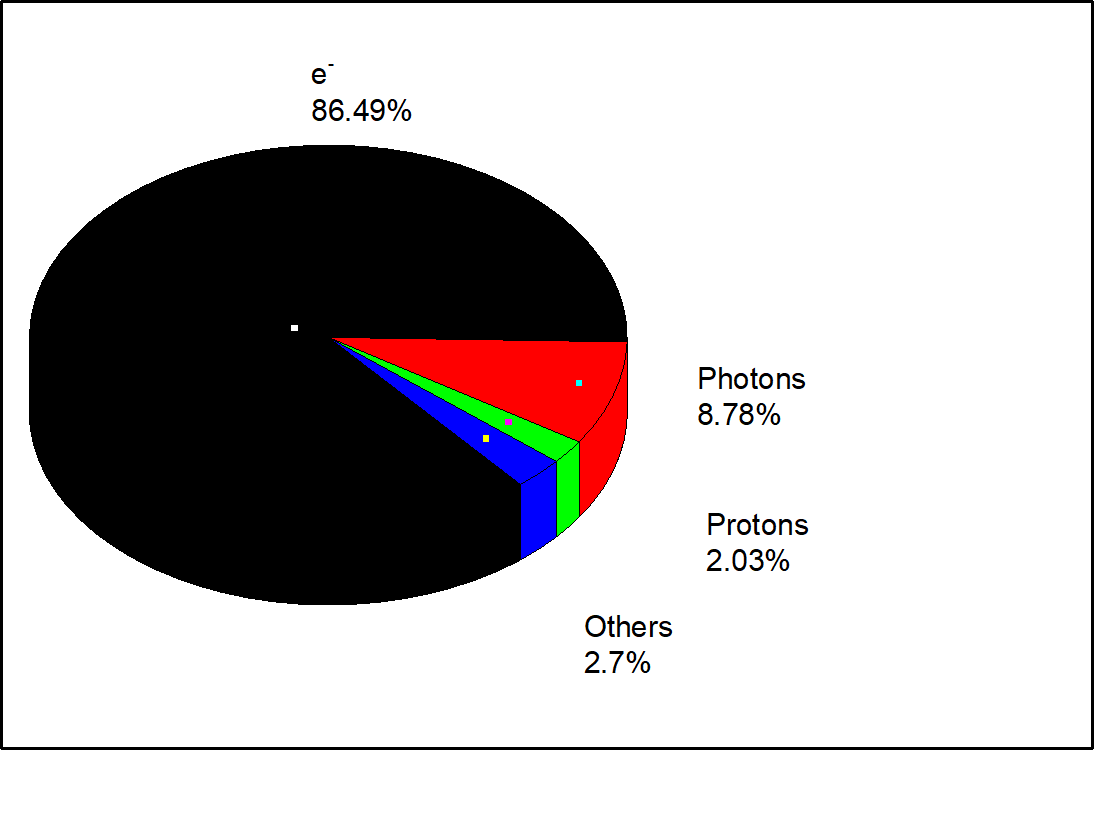}
 \includegraphics[width=0.48\textwidth]{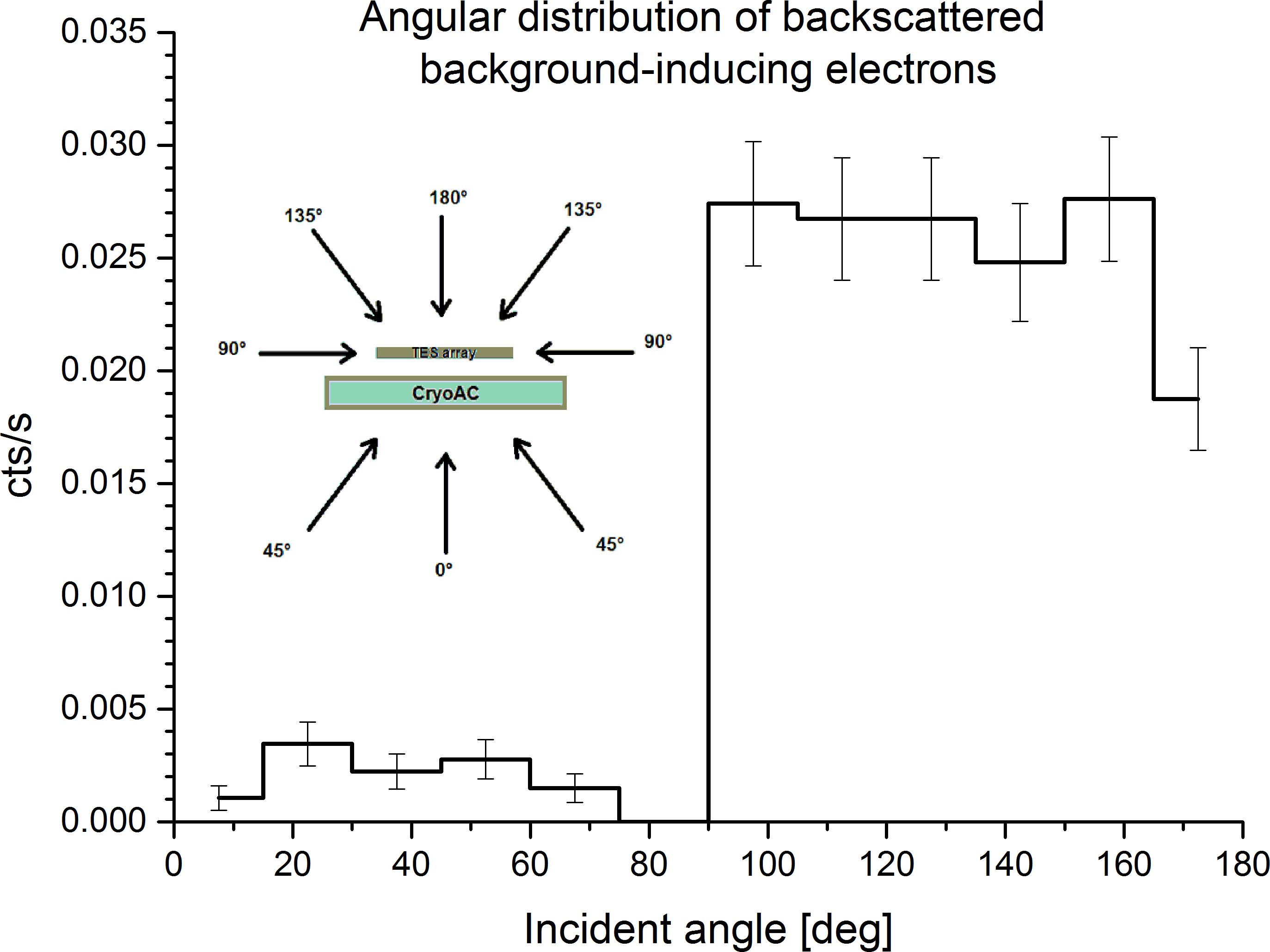}
 \caption{The residual background composition (left), and the angular distribution of incident directions of the backscattered electrons component (right). The plots were produced using GCR protons as input.}
 \label{fig:compdir}
 \end{figure*}

The analysis of the residual background composition and origin revealed that this background is induced mainly by secondary electrons coming from the niobium shield surrounding the detector that impact the X-IFU from above and backscatter on its surface, releasing a small fraction of their energy (see Figure \ref{fig:compdir}). To reduce this component, we tested the secondary electrons yield from different materials, and designed a passive low-Z shield to interpose between the detector and the Niobium. This Kapton shield introduction allowed to further damp the residual background level induced by GCR protons down to $(3.6 \pm 0.16) \times 10^{-3}$ cts $cm^{-2}s^{-1}keV^{-1}$. 

After the Kapton insertion, the background showed a significant contribution ($\sim10\%$) from 16 and 18 keV Niobium fluorescence lines that induce escape peaks on the detector. To prevent these fluorescence photons from reaching the detector several improvements to the passive shielding were tested \citep{lotti2017}, and finally the best option for the passive shield was found to be 10 $\mu$m of Bismuth followed by 250 $\mu$m of Kapton, that allowed to reach the level of $(2.96 \pm 0.16) \times 10^{-3}$ cts $cm^{-2}s^{-1}keV^{-1}$. 

These background levels were obtained applying different screening strategies:
\begin{itemize}
 \item Time coincidence with the CryoAC: photons do not produce a coincidence signal in the X-IFU-CryoAC system, so we can assume that every event that is detected simultaneously ($\Delta t < 10$ $\mu s$) in both the CryoAC and the main detector can be rejected as particle-induced one. A CryoAC low energy threshold of 20 keV is considered.
 \item Pattern recognition: due to the detector features (no charge cloud diffusion among different pixels as in CCD-like detectors, pixels physically separated) source photons will not produce complicate pixel patterns to reconstruct. Instead we can assume that all the events that turn on more than one pixel can be rejected as induced by particles with skew trajectories intersecting more than one pixel, or simultaneous impacts by multiple particles associated to the same primary.
 \item Energy deposition: given the steep decrease of the optics effective area outside the detector energy band, we can assume that all the events that deposit energies outside the instrument sensitivity band are due to charged particles and, therefore, reject them.
\end{itemize}

We performed an investigation on the geometrical origin of the unrejected background. Before the shield introduction the major fraction of the background is induced by the 50 mK Niobium shield, being this the last surface directly seen by the detector. After the Kapton/Bi insertion the contribution from this surface is now mostly blocked by the passive shield, going from ~80\% contribution to 12\%. Even though now the passive shield became the relative main background source, due to a lower electron production yield the absolute background level has been reduced. The corresponding spectra and background level have been reported in Figure \ref{fig:bkglevels}.

 
 
Regarding the residual background particle composition, backscattered electrons produced in the surfaces directly seen by the detector have repeatedly proven to be the main component of the residual background after the insertion of the CryoAC \citep{lotti2018spie,lotti2017}, since $\sim85\%$ of the residual background is induced by secondary electrons, and $\sim80\%$ of these electrons are backscattered. Moreover, we investigated the angular distribution of the backscattered electrons. These particles impact on the detector surface and bounce back, releasing a small fraction of their energy (dependent on the impact angle). As it can be seen from Figure \ref{fig:compdir} (right), almost no electrons backscatter on the lower side of the detector, as the CryoAC efficiently blocks/vetoes those particles, while the major contribution comes from above, as expected. The distribution itself is quite flat, indicating no preferential direction from where these electrons come from, aside from a dip around the direction normal to the detector surface. An analysis of the FPA geometry reveals that this corresponds roughly to the FoV opening angle, the direction where there is almost no mass to produce the secondary electrons.

\begin{figure*}
 \centering
 \includegraphics[width=0.7\textwidth]{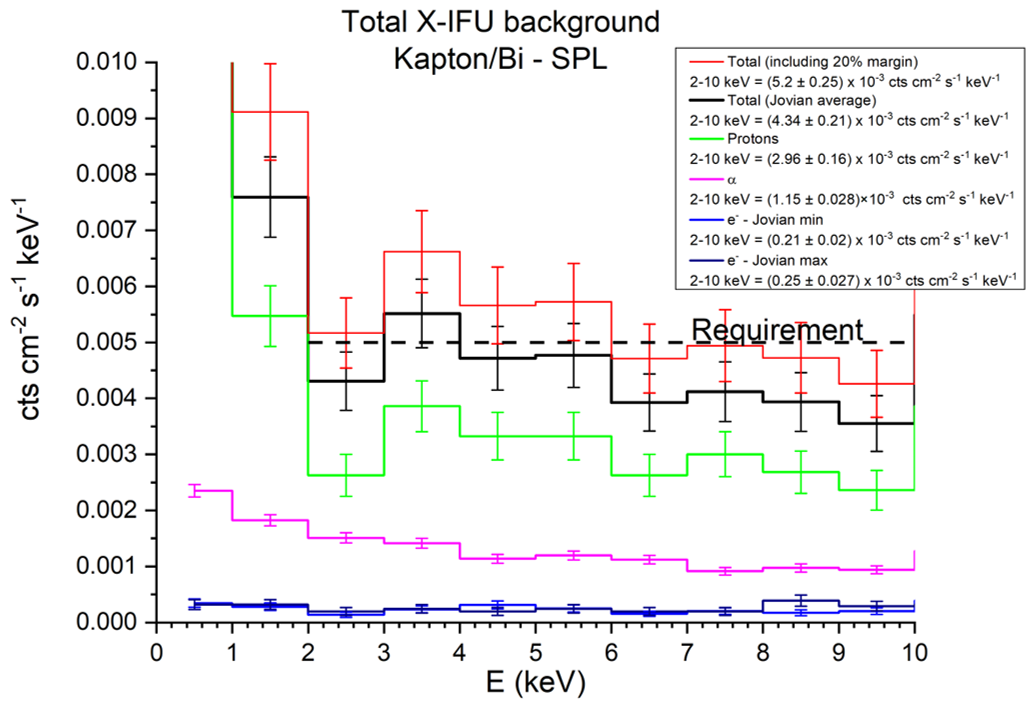}
 \caption{The unrejected particle background of the X-IFU instrument (black line), and all its contributions: GCR protons (red line), alpha (magenta line) and electrons (blue line).}
 \label{fig:totbkg}
 \end{figure*}

Using the updated mass model in the baseline configuration (Kapton/Bi passive shield) we investigated the impact on the background of the other environmental particle populations expected to be present in L2, namely GCR alphas and electrons. 
Our analysis reported a total background level of $(4.34 \pm 0.21) \times 10^{-3}$ cts $cm^{-2}s^{-1}keV^{-1}$ in the 2-10 keV energy band, compliant with the scientific requirement, as it can be seen in Figure \ref{fig:totbkg}. Even if we take into account the 20\% margin the foreseen background level is still compliant with the requirement at 1 $\sigma$ level.

\subsection{Expected variations during mission lifetime}\label{sec:4.2}

\begin{figure*}
 \centering
 \includegraphics[width=\textwidth]{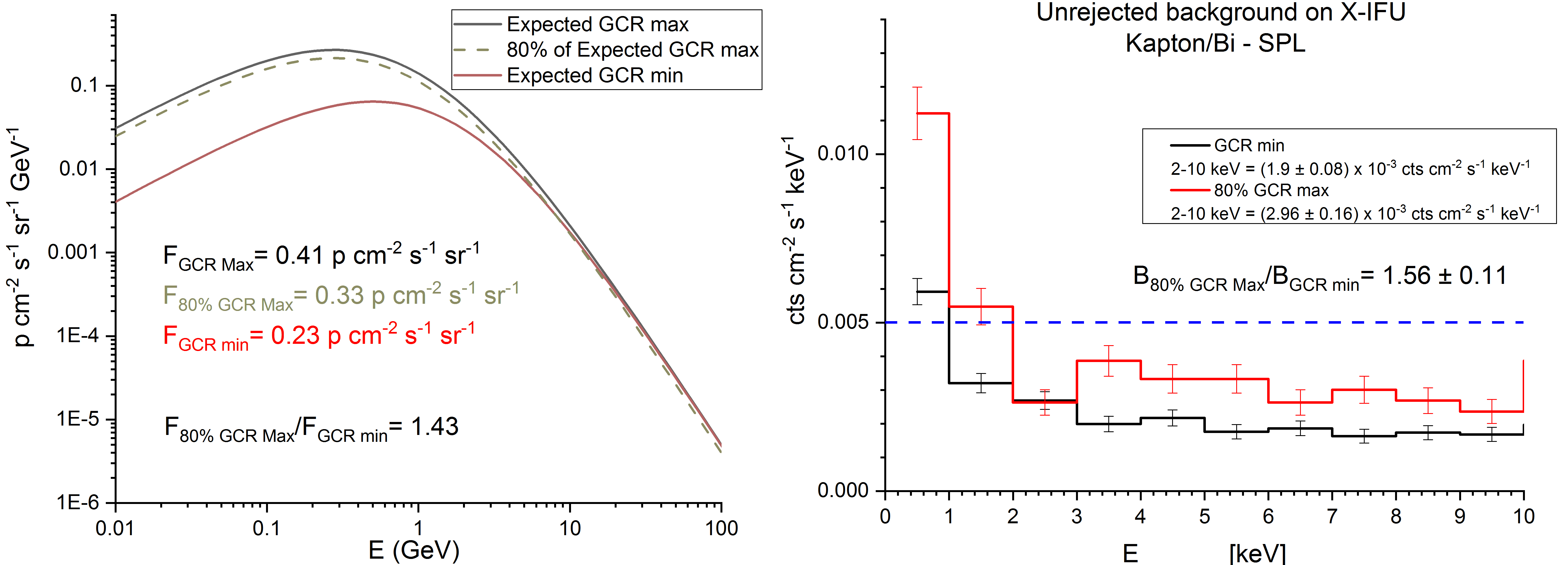}
 \caption{Expected input spectra for the GCR maximum and minimum during Athena mission lifetime (left). Unrejected background spectra resulting from the input spectra on the left (right). Integrated fluxes refer to the 100 MeV - 100 GeV energy band. The residual background scale linearly with the integrated input fluxes.}
 \label{fig:changes}
 \end{figure*}

We investigated the expected variations in the unrejected background level during the mission lifetime, as the GCR flux intensity is anticorrelated with solar phases. We simulated both the maximum and minimum GCR spectra measured in solar cycle 24 after the 2009 year (years 2009 and 2014 respectively, Figure \ref{fig:changes} - left) on the X-IFU mass model, and obtained the results shown in Figure \ref{fig:changes} (right). 

The spectral shape of the unrejected background in the 2-10 keV energy band was not affected sensibly by the change in spectral shape, and the integrated value scaled linearly with the integrated fluxes in the 100 MeV - 100 GeV energy band. 

Regarding the Jovian component of the GCR electrons spectrum, which is expected to vary accordingly to the relative position of Earth and Jupiter during the mission, we expect no significant variation in the induced background spectrum (Figure \ref{fig:totbkg}).

\subsection{Future updates} \label{sec:4.3}

The reported results are based on an X-IFU mass model developed from CAD delivered by CNES and SRON on 2017. The FPA design is constantly under improvement, and since then it has been slightly modified. To establish the impact of such modifications a sensitivity analysis has been performed on some items (see Appendix \ref{sec:sensitivity} and Figures \ref{fig:biau} and \ref{fig:coverage}). The main present differences with respect to the current design are: 

\begin{itemize}
 \item the necessity to change the Bismuth in Gold as a second layer of the passive liner (Kapton based)
 \item The radiation filters modeling needs to be updated to the latest design, in particular the 50 mK filter will be placed at the bottom of the Nb shield and not at the top 
 \item the lower edge of the passive liner must be at a minimum distance higher than 2 mm 
 \item the CryoAC rim needs to be updated, changing the upper 500 um from copper to silicon, according to the latest development
 \item the detector pixel layout needs to be updated to the latest design
 \end{itemize}

furthermore, the simulation settings and post processing can be improved with:

 \begin{itemize}
 \item new settings of the CryoAC detector to improve simulation performances
 \item post processing of the output will be redefined to mimic the actual detector pipeline
\end{itemize}

Most of the bullets have a direct impact onto the FPA design. Thanks to the sensitivity analysis performed, we have solutions to mitigate the impact of these differences to the residual particle background: it is again the use of a well-shaped Kapton/gold liner both around the 50 mK filter metallic frame, and on the inner cylindrical surface of the Nb shield. A new set of Geant4 simulation will be performed once the FPA design becomes more mature. 

\section{The background foreseen for the X-IFU: the low energy component}\label{sec:5}

Soft protons in the $\sim 100$ keV energy range, are funnelled by the X-ray optics towards the focal plane, and have proven to be a major hindrance to the sensitivity of X-ray missions, decreasing by about 40-50\% the usability of the data for background sensitive observations
, and contaminating the remaining data with a poorly reproducible background component (see \citet{lotti2018} and references therein).

The challenging scientific goals of Athena, such as the observation of faint/diffuse/distant sources, impose that soft protons induced background can contribute up to a maximum of 10\% of the high energy particle background requirement (that corresponds to a flux of $5 \times 10^{-4}$ cts $cm^{-2}s^{-1}keV^{-1}$ in the 2-10 keV energy band). The solution is to introduce a high magnetic field between the mirrors and the detectors to deflect these particles away from the instruments field of view (FoV).

\subsection{Mirrors funnelling and filters transmission efficiency}\label{sec:5.1}

The optics funneling efficiency has been calculated by ray-tracing simulations and crosschecked with Geant4 simulations, while the filter transmission efficiency has been characterized by the X-IFU team using a series of mono-energetic simulations (see Figure \ref{fig:L12fluxes} left). This information has been used to construct a protons response matrix, analogous to the one used for photons. Details on the adopted approach are reported in \citet{lotti2018}.

\subsection{The expected fluxes at focal plane level}\label{sec:5.2}

\begin{figure*}
 \centering
 \includegraphics[width=0.48\textwidth]{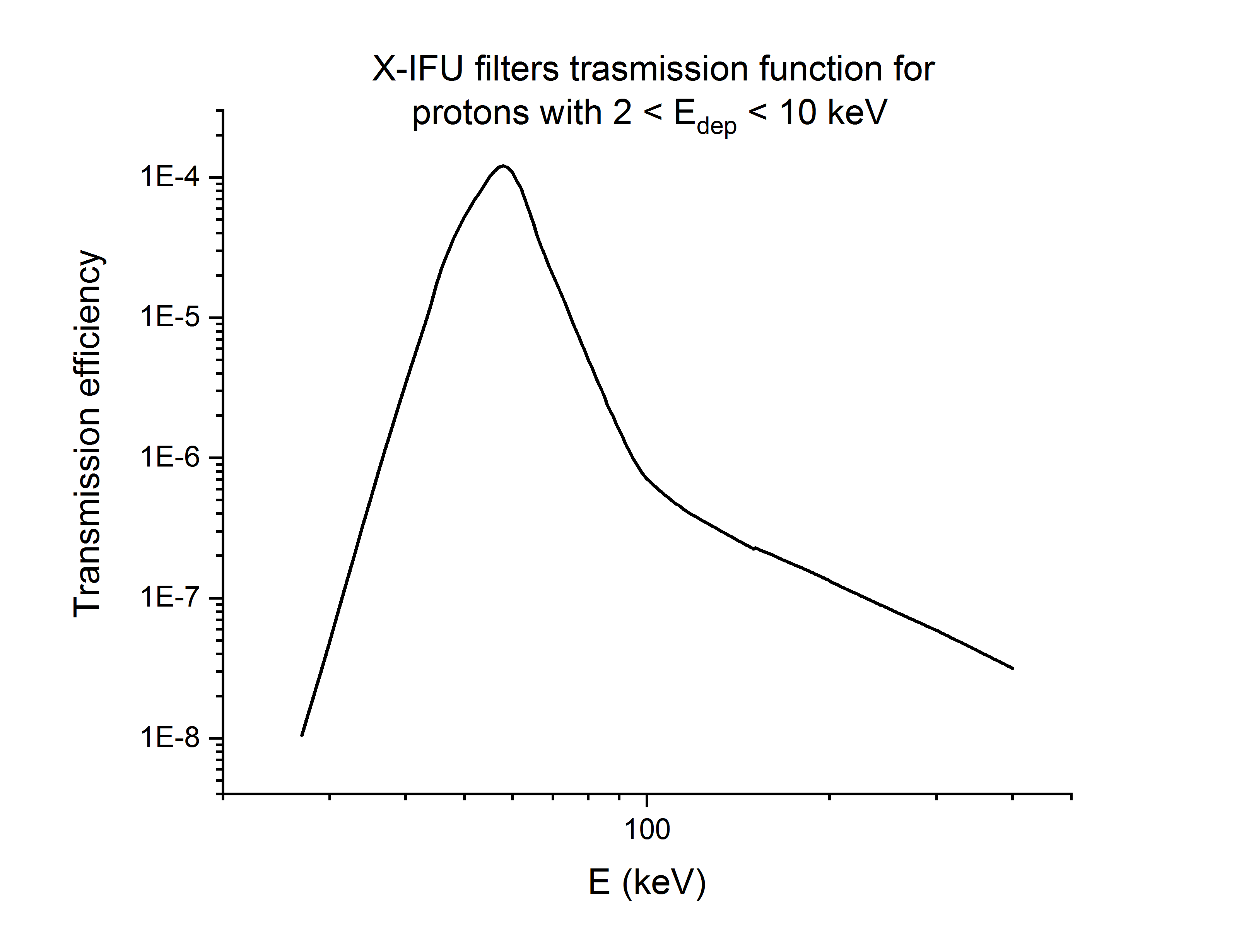}
 \includegraphics[width=0.48\textwidth]{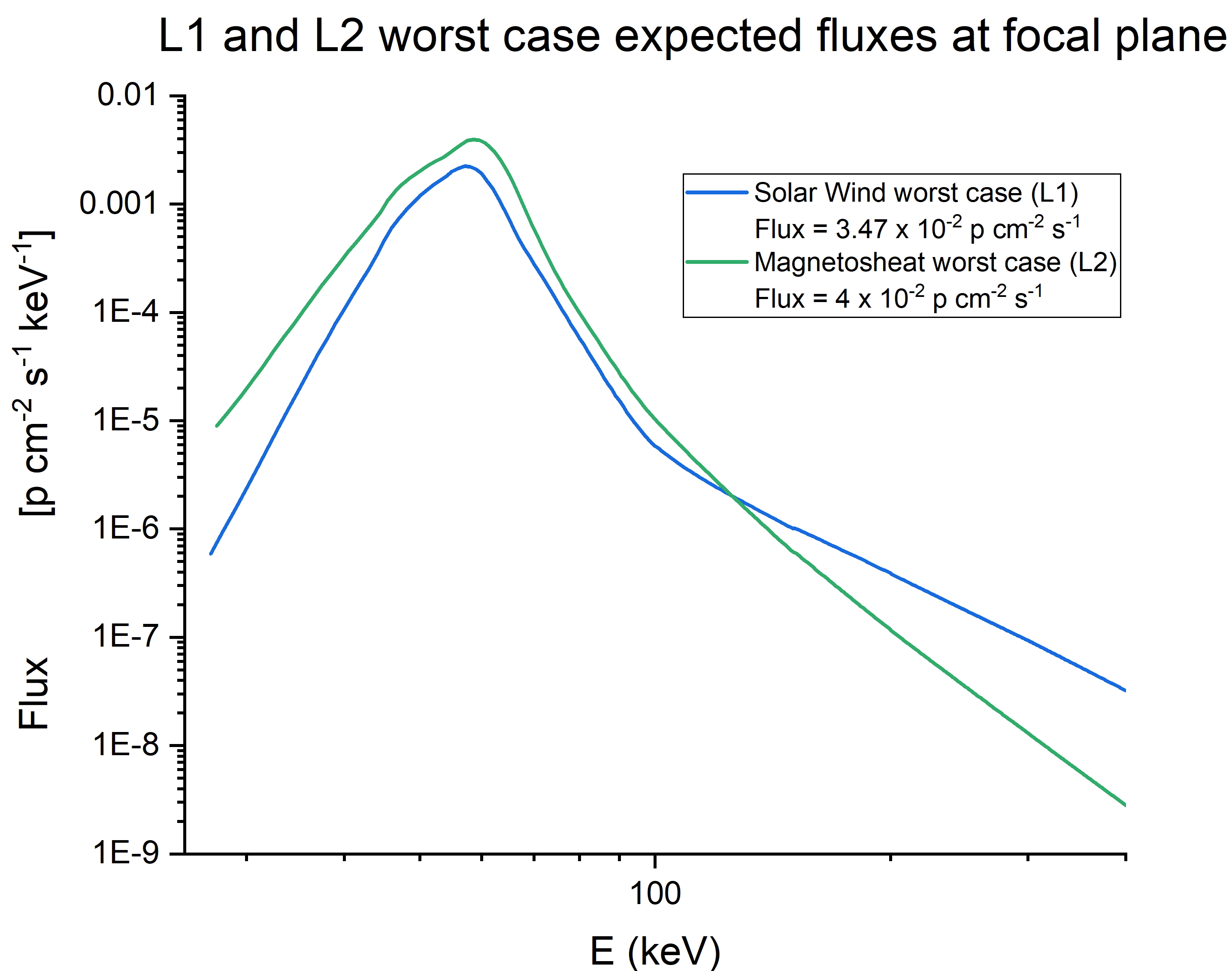}
 \caption{Filters transmission efficiency for Soft Protons for X-IFU, as calculated with Geant4 simulations (left). Expected spectra of the initial energies of the protons that induce background in the 2-10 keV energy range in L1, for the worst case in L1 and the worst L2 case of active Magnetosheath (right), assuming no magnetic diverter.}
 \label{fig:L12fluxes}
 \end{figure*}

The response matrices and the external fluxes have been used to calculate the expected fluxes and spectra of initial energies of the protons that induce background in the 2-10 keV band at focal plane level, as can be seen in Figure \ref{fig:L12fluxes} for L1 and the L2 worst cases \citep{lotti2018}. 
The initial energy spectral distribution of the particles inducing background on the X-IFU is shaped mostly by the transmission of the filters rather than the differences in the slopes of the input spectra.

\begin{table*}\centering
 \caption{Summary table of the relevant numbers. The fractions and energies to block for the 50\% Solar Wind case are not reported since the corresponding fluxes at FPA level already satisfy the requirement.}
 \label{tab:SPresume}
\begin{tabular}{|c|c|c|c|c|}
 \hline
 & Integrated input flux (40-80 keV) & Flux at FPA level &Fraction to block  &E to block\\
 &p $cm^{-2}s^{-1}sr^{-1}$& p $cm^{-2}s^{-1}$  &  &keV\\
 	 \hline
Solar Wind 50\% (L1) & 18.5 & $8.3\times 10^{-4}$ & -  &- \\
Solar Wind 90\%	(L1) & 183.8 & $8.46\times 10^{-3}$ &52.72\% &57\\
Solar Wind worst case (L1) & 751.1 & $3.47\times 10^{-2}$ &  88.47\% & 65\\
Quiet Magnetosheath (L2) & 255.7 & $10^{-2}$ & 60.00\%& 58\\
 Magnetosheath worst case (L2) & 1054.6 & $4\times 10^{-2}$ &	90.00\%	&66\\
 \hline

\end{tabular}
\end{table*}

The integrated background levels for all the cases considered are reported in Table \ref{tab:SPresume}.
Comparing the integrated background levels to the requirement ($5 \times 10^{-4}$ cts $cm^{-2}s^{-1}keV^{-1}$ or, equivalently, $4 \times 10^{-3}$ cts $cm^{-2}s^{-1}$) we can calculate the required fraction of the flux to be blocked by the foreseen magnetic diverter, and from the cumulative curve of the spectral distributions the required energy threshold of such diverter, reported in Table \ref{tab:SPresume}.

The analysis did not show dramatically different flux levels when comparing the two orbits, even though the L1 orbit seems to be favorable in every condition. Consequently, there is not a huge impact of the orbit on the threshold energy to be blocked by the magnetic diverter. 

We remark once again that the coverage of the L1 data is extremely higher than the one relative to L2 ones, since the latter were obtained during just ~2 years of data close to L2 point, i.e. only covering a particular fraction of a single solar cycle. Conversely, we have tens of years of data accumulated by the several satellites to characterize the L1 point. 

\section{Impact of the background on the observation of galaxy clusters with X-IFU}\label{sec:6}

One of the primary goals of Athena X-IFU is probing the diffuse plasma which fills the potential wells of galaxy clusters (the intracluster medium, ICM) \citep{ettori2013hot,pointecouteau2013hot}. For this reason, substantial efforts have been spent in assessing the capabilities of the next-generation Athena
observatory in the characterization of the ICM \citep{cucchetti, cucchetti2, Roncarelli}.

In this section we investigate how well thermodynamic properties of cluster outskirts can be characterized with the X-IFU instrument. We have used XSPEC 12.10.1 \citep{xspec} to simulate typical observations of galaxy clusters at different radii, as foreseen by the Athena Mock Observation Plan, and see how well the instrument can recover the parameters of interest. 

In particular, we have simulated the observation of a mock galaxy cluster at redshift z = 0.1, having a mass $M_{200} = 10^{15} M_{\odot}$ and a characteristic radius $R_{200} = 2$Mpc \footnote{$R_{200}$ is defined as the radius within which the mean density is 200 times the critical density of the Universe, $M_{200}$ is the total mass enclosed within $R_{200}$.}, corresponding to an angular size of $\sim$ 18 arcmin (here we assume a $\Lambda$CDM cosmology with $\Omega_m$ = 0.3, $\Omega_\Lambda$ = 0.7 and H$_0$ = 70 km s$^{-1}$ Mpc). We have modeled the cluster spectrum using the \textit{apec} model \citep{apec} to simulate the emission of the intra-cluster medium, and a \textit{phabs} model \citep{phabs} with N$_H$ = 0.05 $\cdot$ 10$^{22}$ cm$^{-2}$ for the Galactic H absorption. Following \citet{ghirardini2019}, we have assumed for the cluster a temperature profile decreasing steadily with radius, with a logarithmic slope of $\sim -0.3$ in the radial range of interest. The surface brightness at different radii follows measures described in \citet{eckert2012}, while the metal abundances (the elements included are C, N, O, Ne, Mg, Si, S, Ar, Ca, Fe, Ni; He fixed at cosmic; see \citet{andersgrevesse}) have been fixed at the value Z = 0.3 Z$_{\odot}$ \citep{molendi2016}. The mock cluster parameters at different radii are summarized in Table \ref{tab:xspecpar}. 

\begin{table*}\centering
 \caption{Mock cluster parameters used in the simulations.}
 \label{tab:xspecpar}
\begin{tabular}{c c c c}
 \hline

\noalign{\smallskip}

R & kT & Z & S$_{B}$ 0.5 - 2.0 keV \\
$[$R$_{200}$ $]$ & [keV] & [Z$_\odot$] & [erg/cm$^2$/s/arcmin$^2$] \\
\hline
& & & \\
0.2 & 7.72 & 0.30 & 6.53 $\cdot$ $10^{-14}$ \\
0.4 & 6.72 & 0.30 & 1.21 $\cdot$ $10^{-14}$ \\
0.6 & 5.85 & 0.30 & 3.46 $\cdot$ $10^{-15}$ \\
0.8 & 5.10 & 0.30 & 1.17 $\cdot$ $10^{-15}$ \\
1.0 & 4.44 & 0.30 & 4.39 $\cdot$ $10^{-16}$ \\
1.2 & 3.87 & 0.30 & 1.75 $\cdot$ $10^{-16}$ \\

\end{tabular}
\end{table*}

We have simulated multiple X-IFU observations of the cluster using the full FoV of the instrument (5 arcmin equivalent diameter), centering each observation at a different cluster radius, from 0.2 $R_{200}$ to 1.2 $R_{200}$, with a step of 0.2 $R_{200}$. For each pointing, we have assumed an observation time of 100 ks, for a total integration time of 600 ks. The response matrix of the instrument has been obtained from the official documentation of the X-IFU consortium \citep{respmatrix}, and it refers to the so-called\textit{ Athena Cost-Constrained Configuration}, with an optics diameter corresponding to a peak effective area at 1 keV of $\sim$ 1.4 $m^2$\\

To asses the impact of the background on this kind of observations, we have run the simulations using different particle background levels, namely:

\begin{itemize}
 \item "No CryoAC": the background level the instrument will experience without an anticoincidence detector.
 \item "Best Estimate": the current best estimate for the particle background, i.e., $4.34 \times 10^{-3}$ cts $cm^{-2}s^{-1}keV^{-1}$, assuming all the SP are deflected out of the FoV by the magnetic diverter foreseen by the mission.
 \item "GCR min": the minimum background level foreseen, corresponding to the lowest intensity of the GCRs during the mission lifetime.
\end{itemize}

To simulate the observed spectra in each configuration, both the background components (X-ray and particle background) have been modeled inside XSPEC and summed to the model of the source \citep{leccardimolendi2008, leccardimolendi2008-2}. For the CXB we have assumed the model described in \citet{lotti2014}. For the particle background we used a powerlaw with photon index $\alpha$=0 to describe the flat "Best Estimate" and "GCR min" levels. 
The "No CryoAC" level has been instead modeled as the sum of two powerlaws ($\alpha_1$=0, $\alpha_2$=-1.31) and a large gaussian profile roughly describing the GCR Landau peak in the spectrum ($E_l$ = 6.33, $\sigma$ = 1.93). Note that while the cluster and the X-ray background models have been multiplied for both the redistribution matrix (RMF file) and the effective area curve (ARF file) of the instrument, the particle background component has only been convolved with the RMF matrix, since the background particles do not come from the optics.


\begin{figure*}

 \centering
 \includegraphics[width=0.48\textwidth]{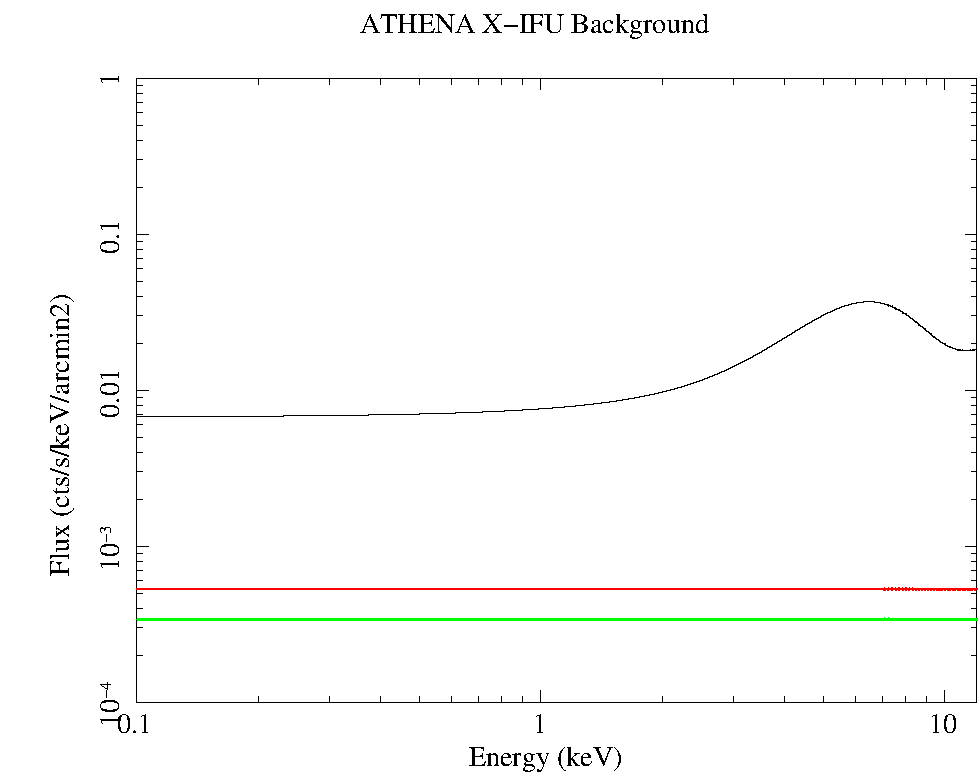}
 \includegraphics[width=0.48\textwidth]{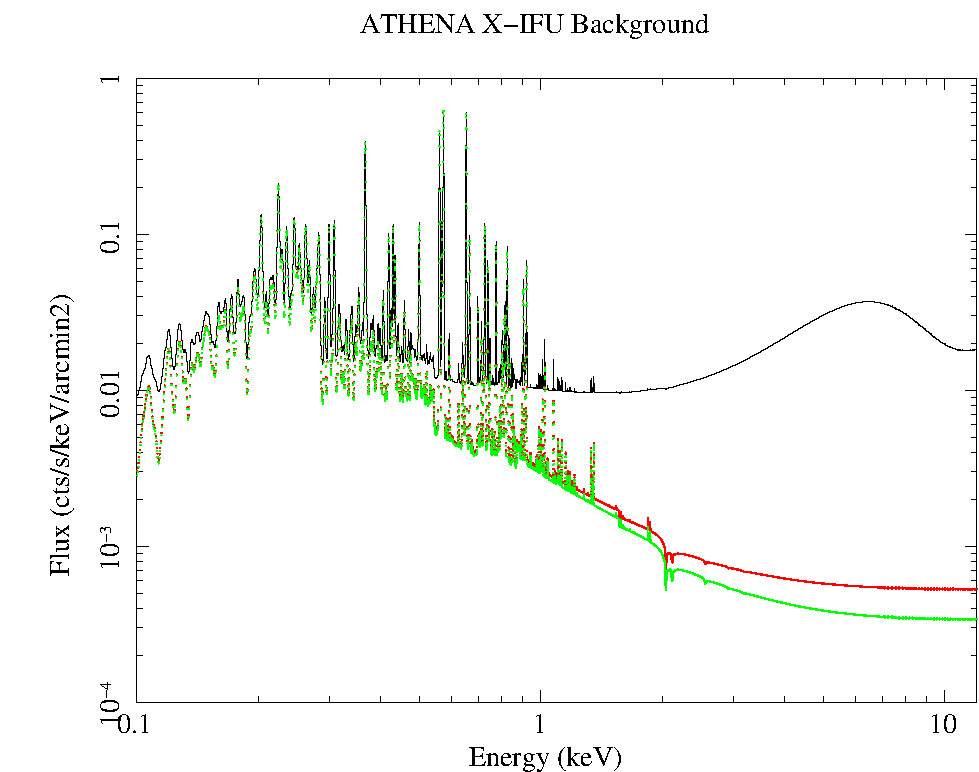}
 \caption{The GCR-induced particle background level considered in the analysis (left), and the total background used in the simulations, including the CXB contribution (right) that starts being dominant below 2 keV. The black line is the ''No CryoAC" background level, the red line is the ''Best Estimate", and the green line is the "GCR min".}
 \label{fig:xspecbkg}
 \end{figure*}

The simulated observations have been generated via the \textit{fakeit} routine, using counting statistics in creating the spectra. The data have been then grouped to have at least 1 count in each spectral bin. Finally, we have performed the spectral fitting to recover the cluster temperature and metallicity, using the Cstat statistic. The fit has been performed by using the 3 components model used to simulate the observation (cluster + Xray background + Particle background). In this procedure, we have constrained the fit in the 0.2-10 keV energy band, and froze the Galactic absorption column nH and the redshift z of the source. For the background components, we fixed all the parameters, except for the normalizations.

\begin{figure*}

 \centering
 \includegraphics[width=0.48\textwidth]{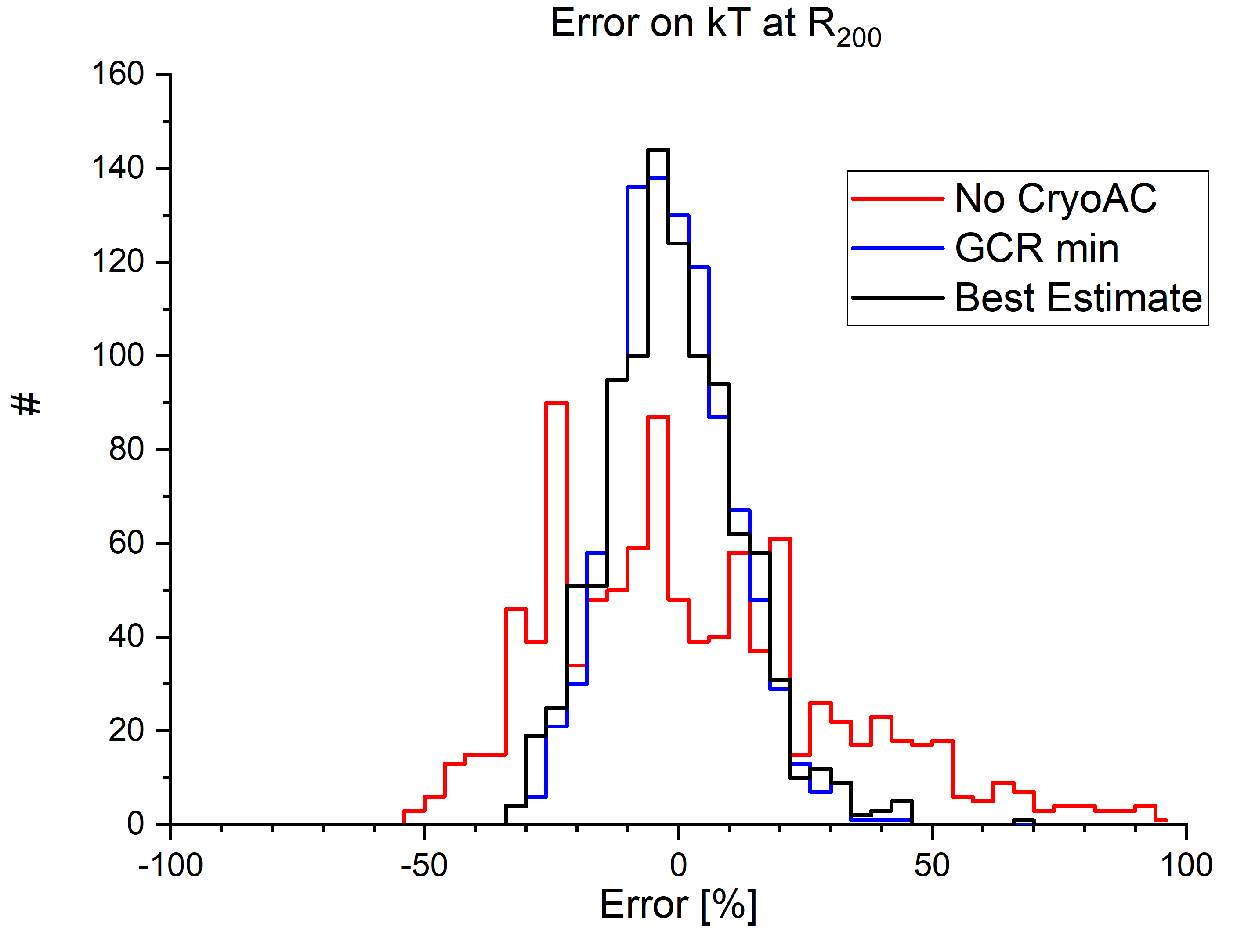}
 \includegraphics[width=0.48\textwidth]{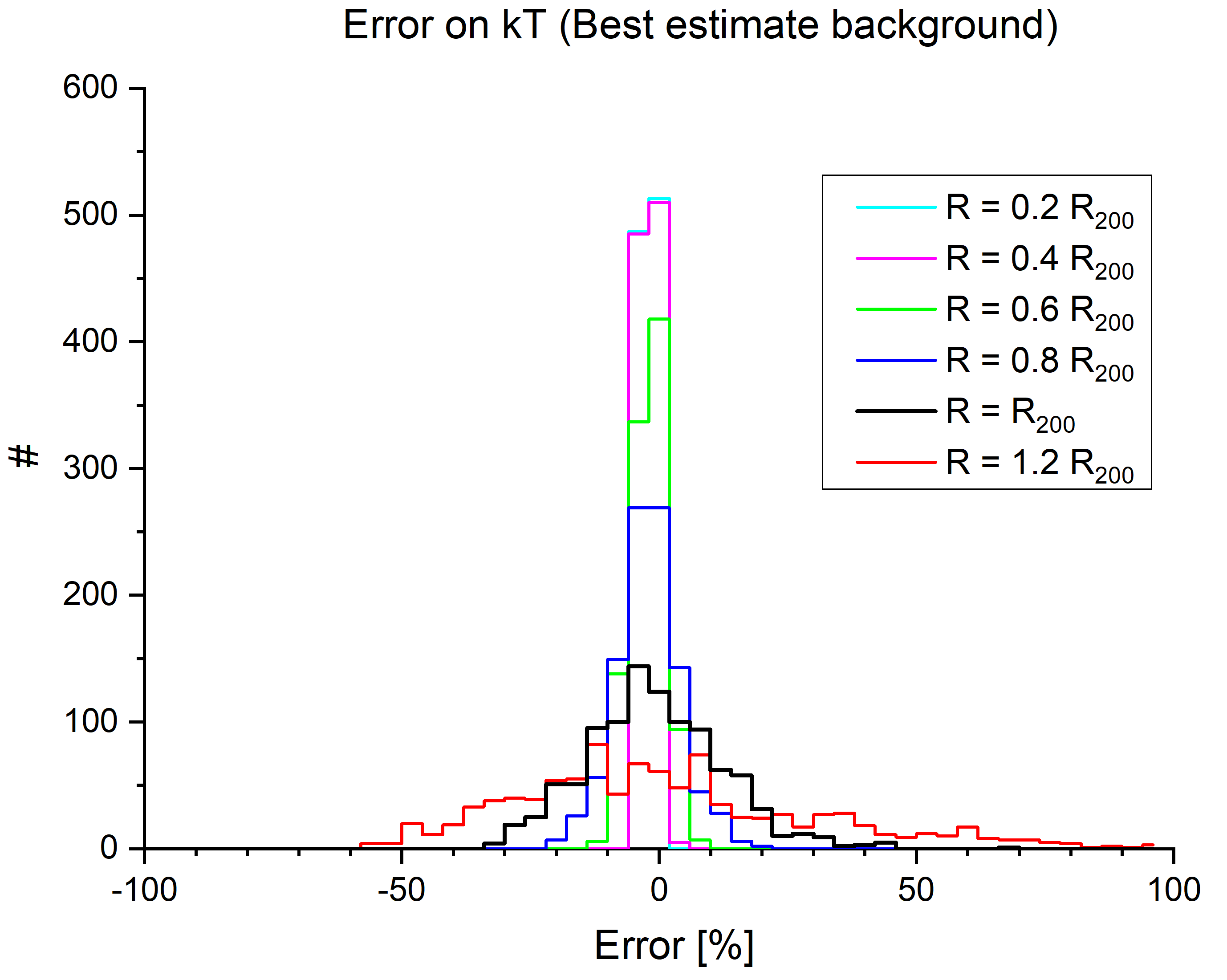}
 \caption{Distributions of the errors in the recovered cluster temperature, referring to 1000 simulated observation per configuration. The Error is defined as (kT$_{REAL}$ - kT$_{RECOVERED}$)/kT$_{REAL}$. Left: Observation at a fixed cluster radius (R$_{200}$) with different background levels (NoCryoAC, Best Estimate, GCR min). Right: Observation at different cluster radii (0.2, 0.4, 0.6, 0.8, 1.0, 1.2 R$_{200}$) with the "Best estimate" background level.}
 \label{fig:clustergaussian}
 \end{figure*}

To investigate with what accuracy it is possible to recover the cluster parameters, and evaluate the statistical uncertainties of the fit results, we have repeated each simulation 1000 times. In Figure \ref{fig:clustergaussian} we report the distributions of the errors in the recovered cluster temperature, for a fixed cluster radius and different background levels (Figure \ref{fig:clustergaussian} left), and for a fixed background level and different cluster radii (Figure \ref{fig:clustergaussian} right). The distributions related to the recovered metallicity show the same trends.

For each distribution we have calculated the mean value and the standard deviation of the recovered parameters. The results of this analysis are reported in Table \ref{tab:clustersim}. 
For both temperature and abundance, we see how measurements can be extended to larger radii as the background intensity decreases from the "No CryoAC" to the "GCR min" case. As expected, the biggest jump in sensitivity is achieved by adding the anti-coincidence. Indeed, without the CryoAC, investigation of clusters would be restricted to the core and circum-core regions. Almost no improvements are registered as we move from the Best Estimate to the GCR min background intensities. Broadly speaking we can say that measures of both temperature and metallicity can be extended to $R_{200}$ if the Best Estimate background level can be achieved.

\begin{figure*}

 \centering
 \includegraphics[width=0.48\textwidth]{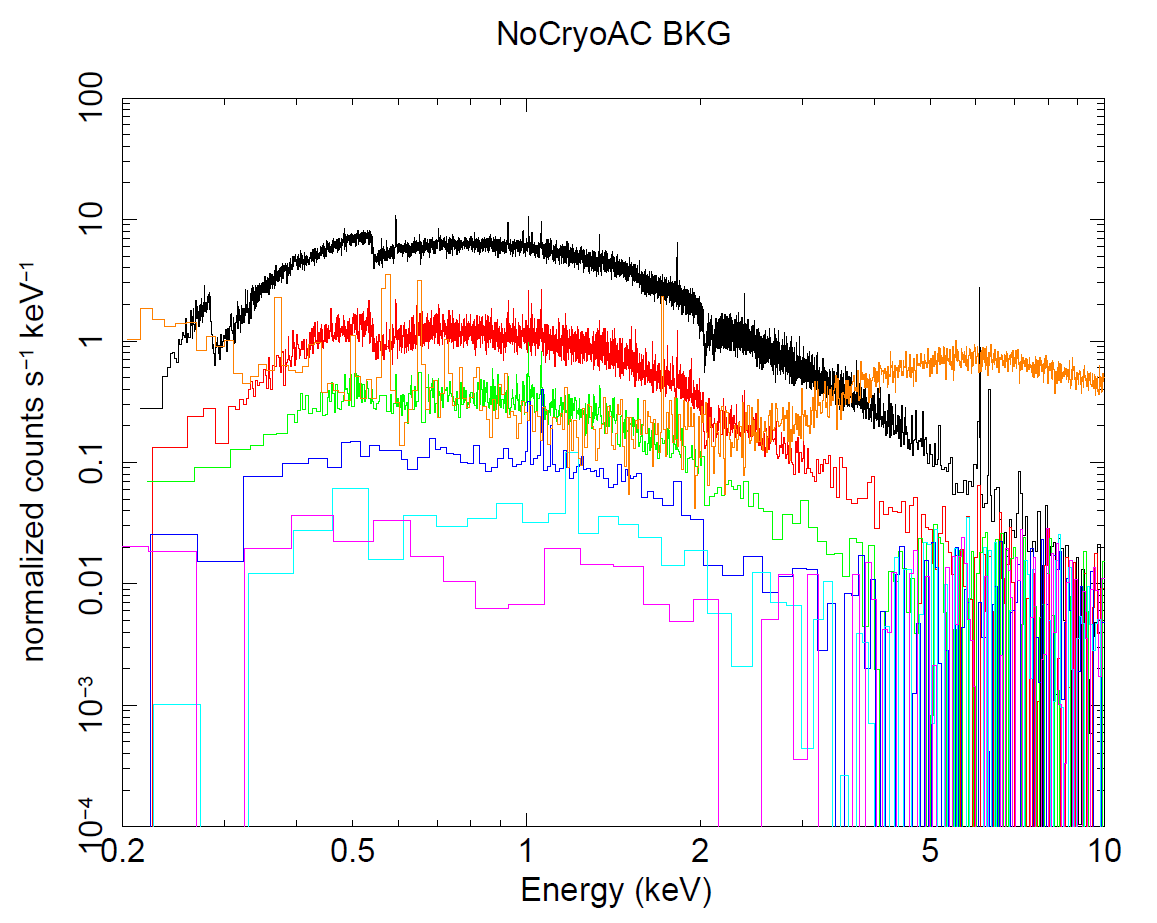}
 \includegraphics[width=0.48\textwidth]{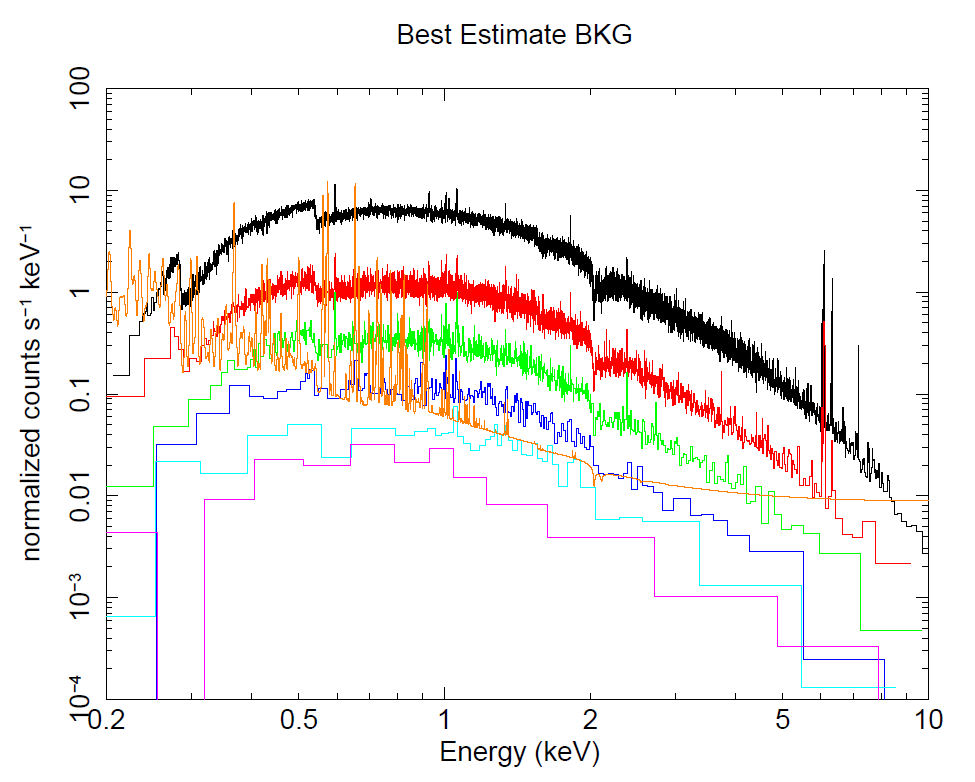}
 \caption{Simulated cluster spectra with the different background configurations. In each plot the black, red, green, blue, cyan and purple lines refer to the 100 ks pointing at 0.2, 0.4, 0.6, 0.8, 1.0, 1.2 R$_{200}$ , respectively. The background levels are overplotted in orange. (Left) : "No CyoAC" backgorund level. (right): "Best Estimate" background level.}
 \label{fig:clustersimpict}
 \end{figure*}

\begin{table*} 
 \caption{Mean value and standard deviation of the cluster parameters recovered fitting 1000 times the simulated spectra with the different background levels. The err columns represent the ratio between the mean value and the $\sigma$.}
 \label{tab:clustersim}
\centering
\begin{tabular}{|c c|c c c|c c c|c c c|c c c|}
 \hline


 \multicolumn{2}{|c|}{Model} & \multicolumn{3}{c|}{GCR min} & \multicolumn{3}{c|}{Best Estimate} & \multicolumn{3}{c|}{No CryoAC} \\


R & kT & kT & $\sigma$ & err & kT & $\sigma$ & err & kT & $\sigma$ & err  \\
 
$[$ $R_{200}$ $]$ & [keV] & [keV] & [keV] & [\%] & [keV] & [keV] & [\%] & [keV] & [keV] & [\%]  \\ 
 
\hline


0.2& 7.72& 7.72	& 0.05	& 1	& 7.72	& 0.05	& 1	& 7.72	& 0.06	& 1	 \\
0.4& 6.72& 6.72	& 0.08	& 1	& 6.72	& 0.08	& 1	& 6.72	& 0.13	& 2	 \\

0.6& 5.85& 5.84	& 0.19	& 3	& 5.84	& 0.20	& 3	& 5.84	& 0.32	& 5		 \\

0.8& 5.10& 5.09	& 0.28	& 6	& 5.10	& 0.31	& 6	& 5.17	& 0.64	& 12	 \\

1.0& 4.44& 4.49	& 0.52	& 12& 4.49	& 0.59	& 13& 4.68	& 1.32	& 28 \\

1.2& 3.87& 4.03	& 1.22	& 30& 4.04	& 1.27	& 31& 4.95	& 4.36	& 88 \\

& & & & & & & & & & \\
\hline

R & Z & Z & $\sigma$ & err & Z & $\sigma$ & err & Z & $\sigma$ & err  \\

$[$ $R_{200}$ $]$ & [Z$_\odot$] & [Z$_\odot$] & [Z$_\odot$] & [\%] & [Z$_\odot$] & [Z$_\odot$] & [\%] & [Z$_\odot$] & [Z$_\odot$] & [\%] \\

\hline

0.2&	0.30&	0.30&	0.01&	3&	0.30&	0.01&	3&	0.30&	0.01&	3\\
0.4&	0.30&	0.30&	0.01&	3&	0.30&	0.01&	3&	0.30&	0.02&	7\\
0.6&	0.30&	0.30&	0.02&	7&	0.30&	0.02&	7&	0.30&	0.04&	13\\
0.8&	0.30&	0.30&	0.04&	13&	0.30&	0.04&	13&	0.32&	0.08&	27\\
1.0&	0.30&	0.31&	0.08&	26&	0.31&	0.08&	26&	0.36&	0.23&	64\\
1.2&	0.30&	0.34&	0.19&	56&	0.35&	0.22&	65&	0.65&	0.94&	144\\

 \hline

\end{tabular}
\end{table*}

    In the analysis described so far, we only dealt with the statistical fluctuations of the background, assuming no systematic uncertainties. To evaluate also the impact of systematics on this kind of studies, we have then repeated the analysis assuming a 2\% uncertainty on the particle background component. This value represents the current requirement for the X-IFU Non-X-ray background level. To do this, we have re-generated the simulated spectra by scaling with a random factor in the range [0.98, 1.02] the normalization of the particle background component (for each one of the 1000 simulation per configuration). We then performed the fit forcing the normalization of the background component to assume the unperturbed value. 
    We do not notice any significant difference with respect to the previous analysis. 
    This seems surprising since, for the larger radii, the source to background ratio around 6 keV is very small and variations of a few percent on the background should have an impact on measurements. The explanation comes from the presence of L-shell emission lines, around 1 keV, that become progressively stronger as we move to larger radii and smaller temperatures. The abundance is measured from these lines which, as can be seen in Figure \ref{fig:clustersimpict}, are only very mildly affected by the background. 
    The role of Fe L-shell emission in the measurement of metal abundances in cluster outskirts is extensively discussed in \citep{ghizzardi2021}. In that paper the authors point out that L-shell abundances derived with the limited spectral resolution of a CCD detector can be, and in the specific case are,  affected by significant biases.
    The X-IFU, with its unprecedented spectral resolution, will not suffer from such issues, however the same authors warn that to achieve a reliable estimate of the metal abundance from the Fe  L-shell lines we need to reach a deeper understanding of the physical processes responsible for the production of the lines. Just to give an example, the lines come from ionization states that, at the relevant temperature, account for no more than 10\% of the total iron, thus collisional ionization equilibrium has to be understood at the few percent level if we are to make reliable use of the data.

\section{Summary and conclusions}\label{sec:7}

The goal of this paper has been to provide the current expectation of the X-IFU residual particle background, and how it affects the typical observation of a faint/diffuse source. In this respect we used Monte Carlo simulations by means of the Geant4 toolkit to assess the particle backgrounds, and the XSPEC software to simulate the astrophysical observation.

In order to increase the reliability of the Monte Carlo simulations for the background estimates, we adopted the most up-to-date models for the L1 and L2 environments and for the instrument. Our update work on the previous version of the mass model showed that the unrejected background level is quite robust with respect to changes that happen far from the detector.

Regarding the high-energy GCR-induced component of the background, we found that using only the main detector array the X-IFU would exceed the background level requirement. For this reason, the X-IFU includes a cryogenic anticoincidence detector (CryoAC), that allows to discriminate high energy particles that cross the detector, reducing the unrejected background level down to $(4.8 \pm 0.16) \times 10^{-3}$ cts $cm^{-2}s^{-1}keV^{-1}$ (Figure \ref{fig:bkglevels}).
This background level is induced mainly by secondary electrons that backscatter on its surface, releasing a small fraction of their energy, for which the CryoAC is useless. The introduction of a low-Z shielding surrounding the detector, and its further optimization (see Section \ref{sec:4}), allowed to reduce the background level induced by GCR protons down to $(2.96 \pm 0.16) \times 10^{-3}$ cts $cm^{-2}s^{-1}keV^{-1}$. Accounting also for the contribution of GCR alphas and electrons, the total unrejected background level is $(4.34 \pm 0.21) \times 10^{-3}$ cts $cm^{-2}s^{-1}keV^{-1}$. 
This background level is compliant with the requirement of $5 \times 10^{-3}$ cts $cm^{-2}s^{-1}keV^{-1}$, enabling the several background-sensitive scientific objectives of the mission. Even if we take into account the 20\% margin the foreseen background level is still compliant with the requirement at 1$\sigma$ level.
Furthermore, we also investigated the initial energies of the particles inducing background on the X-IFU, the effect of specific changes to the FPA configuration, as well as the rejection efficiency of the CryoAC detector as a function of its positioning/sizing. Finally, we calculated the expected variations of the background during the mission lifetime, finding that a factor of $\sim1.5$ reduction is expected, according to the reduction of the external flux with respect to the maximum value adopted here.

Regarding the low-energy induced component, namely the Soft Protons, a response matrix has been constructed and used to calculate the flux at the focal plane level for all the external conditions reported in Section \ref{sec:spectra}. By comparing these fluxes to the requirement of $5 \times 10^{-4}$ cts $cm^{-2}s^{-1}keV^{-1}$ (2-10 keV) we found the fraction that has to be blocked by the insertion of a magnetic diverter, and the corresponding energy up to which the flux has to be damped.
The analysis did not show dramatically different flux levels between the L1 and L2 orbits, being the incident spectra shaped mostly by the filters transmission rather than the initial shape, even though the former seems to be favorable in every condition. Consequently, there is not a huge impact of the orbit on the threshold energy to be blocked by the magnetic diverter. 
However, there is an extreme lack of data in the L2 orbit (obtained only during just ~2 years of data close to L2 point), while the significance of the L1 data is extremely higher (tens of years of data accumulated by the several satellites). Being this a limitation related to the very existence of data related to L2 point rather than their analysis, we can only label the L2 environment as very weakly characterizable in regards to soft protons until new data become available, and advice against its choice as orbit for SP-sensitive missions.

Finally, we used the calculated background levels to simulate the observation of a mock galaxy cluster from its central region, out to $1.2$ $R_{200}$. We simulated 100 ks observations with all the calculated background levels: without the CryoAC,  with the best estimate (GCR max conditions), and with "GCR min" level.

These results provide evidence that without an AntiCoincidence detector it is not possible to characterize the properties of the cluster at large radii, highlighting the essential role of this instrument. The background reduction techniques illustrated in this paper are therefore mandatory to study the characteristics of the hot plasma in the cluster outskirts regions. 
We found no improvement in the observations obtained with the "GCR min" background with respect to the "Best estimate" (GCR max) background, suggesting that if there are no uncertainties in the atomic modeling of the Fe L lines moving cluster outskirt observations to periods with the lowest GCR intensity might not be advantageous.
Finally, thanks to the extreme energy resolution of the X-IFU, even accounting for a 2\% uncertainty on the particle background component did not impact the instrument possibility to recover the parameters of interest.

\section*{Acknowledgements}
The present work has been supported by the ESA AREMBES Core Technology Program (CTP), contract No 4000116655/16/NL/BW.
This work has been supported by ASI (Italian Space Agency) through the Contracts no. 2015-046-R.0, 2018-11-HH.0 and 2019-27-HH.0. We also thank CNES for allowing the use of their computing system to obtain the simulation results in a reasonable timeframe, and SRON for the provided material and useful discussion.
The authors thank the P.I. Didier Barret for the useful discussion.

\section*{Appendix}

\section*{A. Simulation settings and normalization procedure}\label{sec:normalization}

There are several models in the Geant4 toolkit that treat the same physical processes, with different internal settings and applicability conditions. Which models to use, and their internal settings, are defined in the so-called Physics List file. The Geant4 consortium provides several pre-assembled Physics Lists, optimized for different purposes. 
We started the background evaluation activity using the pre-assembled Physics List that according to software developers was the most suited for space-based applications, namely the G4EmStandardPhysics\_option4 (Opt4). 
However, given the number of physical applications Geant4 can be used to simulate, it is impossible to validate the software behavior in every possible situation. So, thanks to ESA support, we defined our own Physics List dedicated to Athena, the “Space Physics List” (SPL)\footnote{The SPL is available for download at the following url: \url{https://file.sic.rm.cnr.it/index.php/s/xsVgQnTtWG5Rbqq}}, comparing the performances of different Geant4 models in reproducing the available experimental data for the physical processes involved in the generation of background. 
At the end of this activity the SPL was officially endorsed by ESA for the X-IFU, and we adopted it as our reference settings for Athena simulations. 

\begin{figure*}
 \centering
 \includegraphics[width=0.48\textwidth]{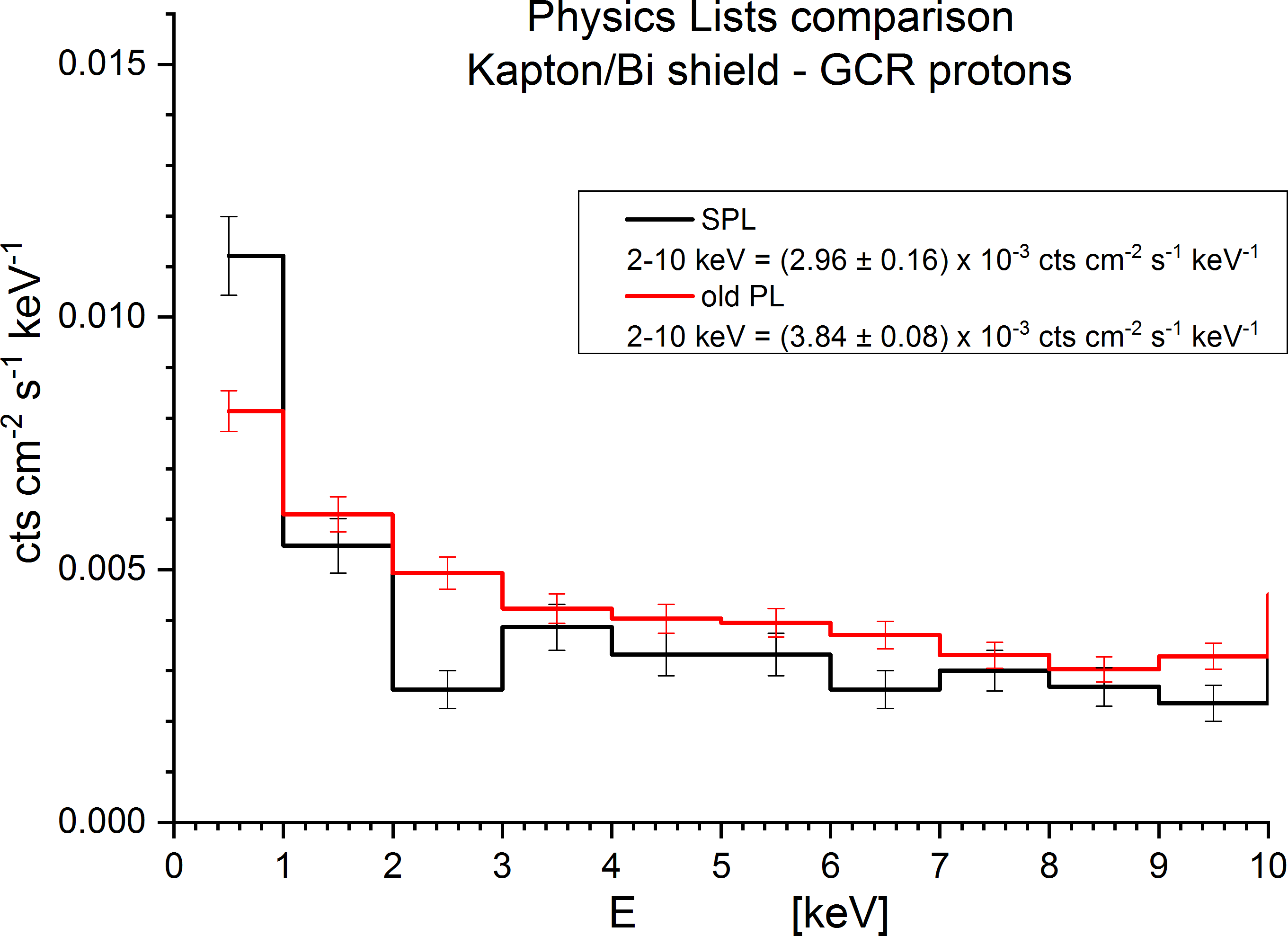}
 \includegraphics[width=0.48\textwidth]{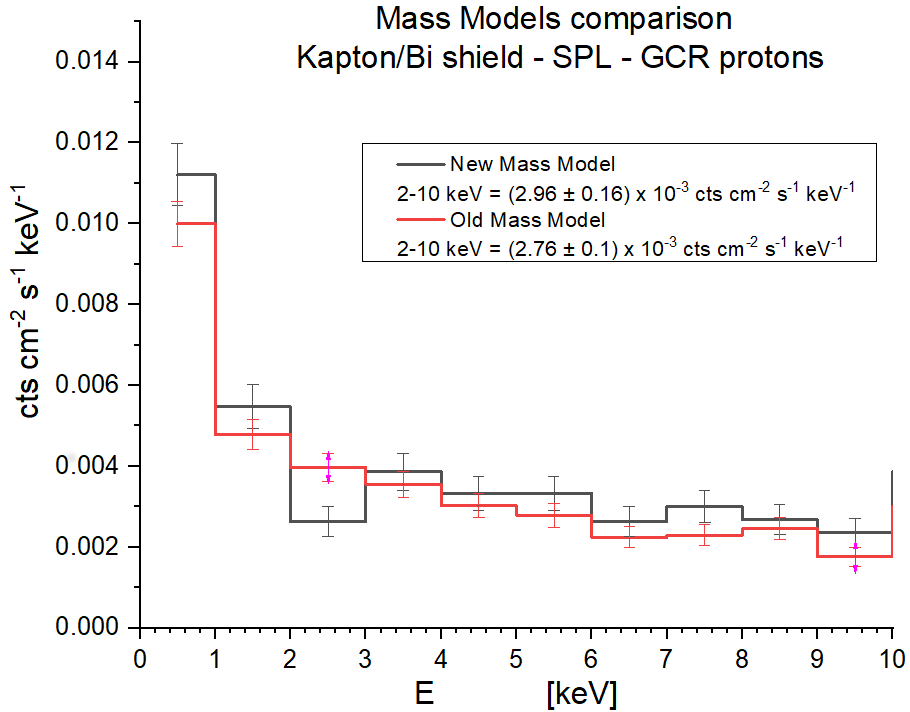}
 \caption{Left: Residual background level obtained with the two physics lists described in the text. Right: The residual background spectra from the GCR protons, obtained with the old and new mass models using the Space Physics List (see Appendix \ref{sec:normalization}).}
 \label{fig:PLMMcomparison}
 \end{figure*}

We then compared the effect of the two physics lists on the simulation results, using as inputs the same GCR protons spectrum (see Section \ref{sec:spectra}) and the same mass model and the 4.10.2 version of the Geant4 software. The results are shown in figure \ref{fig:PLMMcomparison}, and show that there is a $\sim20\%$ reduction in background rates when we go from the old to the new settings.
 
In the standard configuration the different solids in the mass model are assigned to different regions within Geant4, each one having different settings of the cutoff for the generation of secondary particles:

\begin{itemize}
 \item 	The detector, the supports, and the surfaces directly seen by the detector are assigned to the "inner region" with the lowest possible cut values (few tens of nm, high detail level)
 \item 	The remaining solids in the FPA are assigned to an "intermediate region" with higher cut values (few $\mu m$)
 \item 	The cryostat and the masses outside the FPA are assigned to the "external region" where the cut (few cm) allowed the creation only of high energy secondary particles
\end{itemize}

we remind that in Geant4 the cutoff specifies the energy threshold (in terms of traveled distance) below which the secondary particles are not produced, and the regions are subsets of the mass model where specific settings (such as the cutoff, or specific physical models) can be used.

We tested also the dependence of the background on the cutoff value of the different regions. Regarding the inner region, we found out that increasing the cut for gammas, $e^-$, and $e^+$ from 0.05 $\mu m$ to 0.5 $\mu m$ allows a decrease of the total CPU time of about 60\% without losing precision in the background simulation. The use of the Space Physics List, without the single scattering, gives consistent results, within the errors, with the rates obtained using the reference Opt4 physics list, while increasing the CPU time of about 30-40\%. 
 
\begin{figure*}
 \centering
 \includegraphics[width=0.5\textwidth]{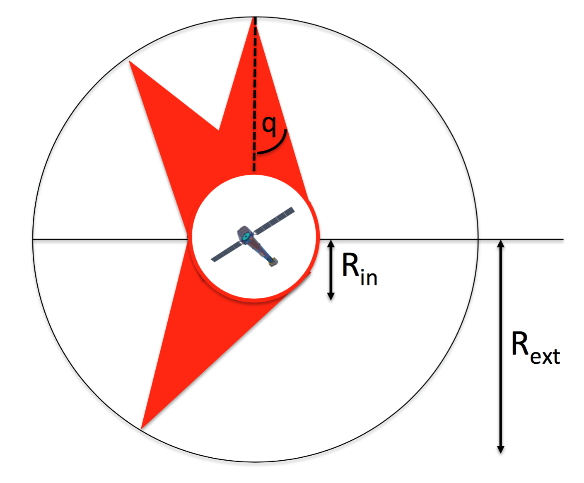}
 \caption{Schematic view of the angular and geometrical distribution of the input particles. $R_{in}$ is the radius surrounding the mass model, $R_{ext}$ the radius from which particles are shot.}
 \label{fig:normalization}
 \end{figure*}

The normalization procedure of the simulations, namely the process that leads from the number N of emitted particles to a particle or count rate, has been reported in \citet{normalization3}. The simulated exposure time T (in seconds) to the isotropic flux depends on the number of simulated particles N as follows:

 \begin{equation}\centering
 T=\frac{N}{\phi \times 4\pi^2R_{ext}^2 \times sin^2 q}~~~~~~~[\textrm{seconds}]
 \end{equation}

where we assume that the particles are emitted from a spherical surface of radius $R_{ext}$ within a cone of half angle (q), following a cosine law angular distribution (see Figure \ref{fig:normalization}), and where $\phi$ is the energy integrated particle intensity in space, in units of particles $cm^{-2} s^{-1} sr^{-1}$.

\section*{B. Sensitivity analisys} \label{sec:sensitivity}

 \begin{figure*}
 \centering
 \includegraphics[width=0.48\textwidth]{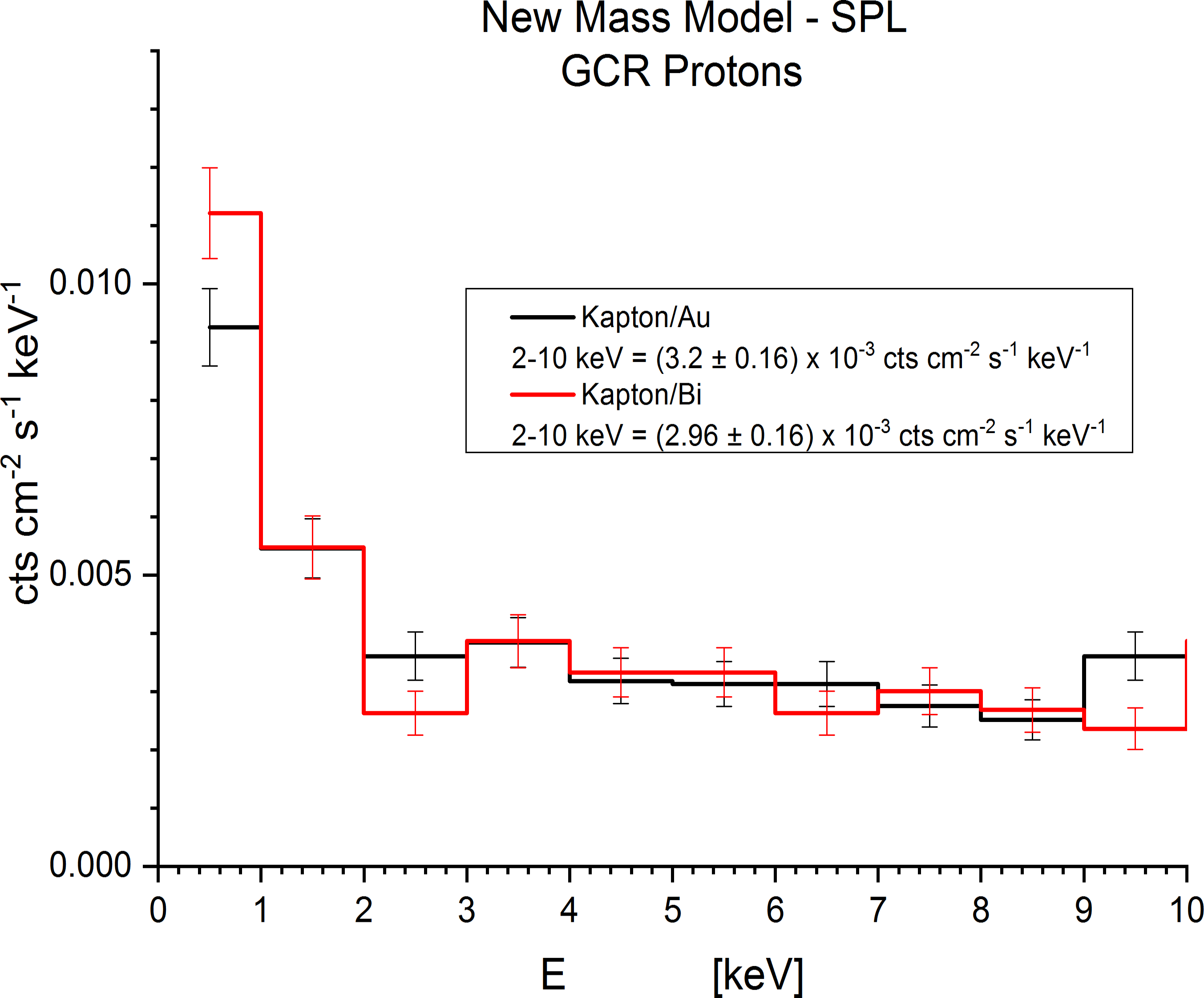} \includegraphics[width=0.48\textwidth]{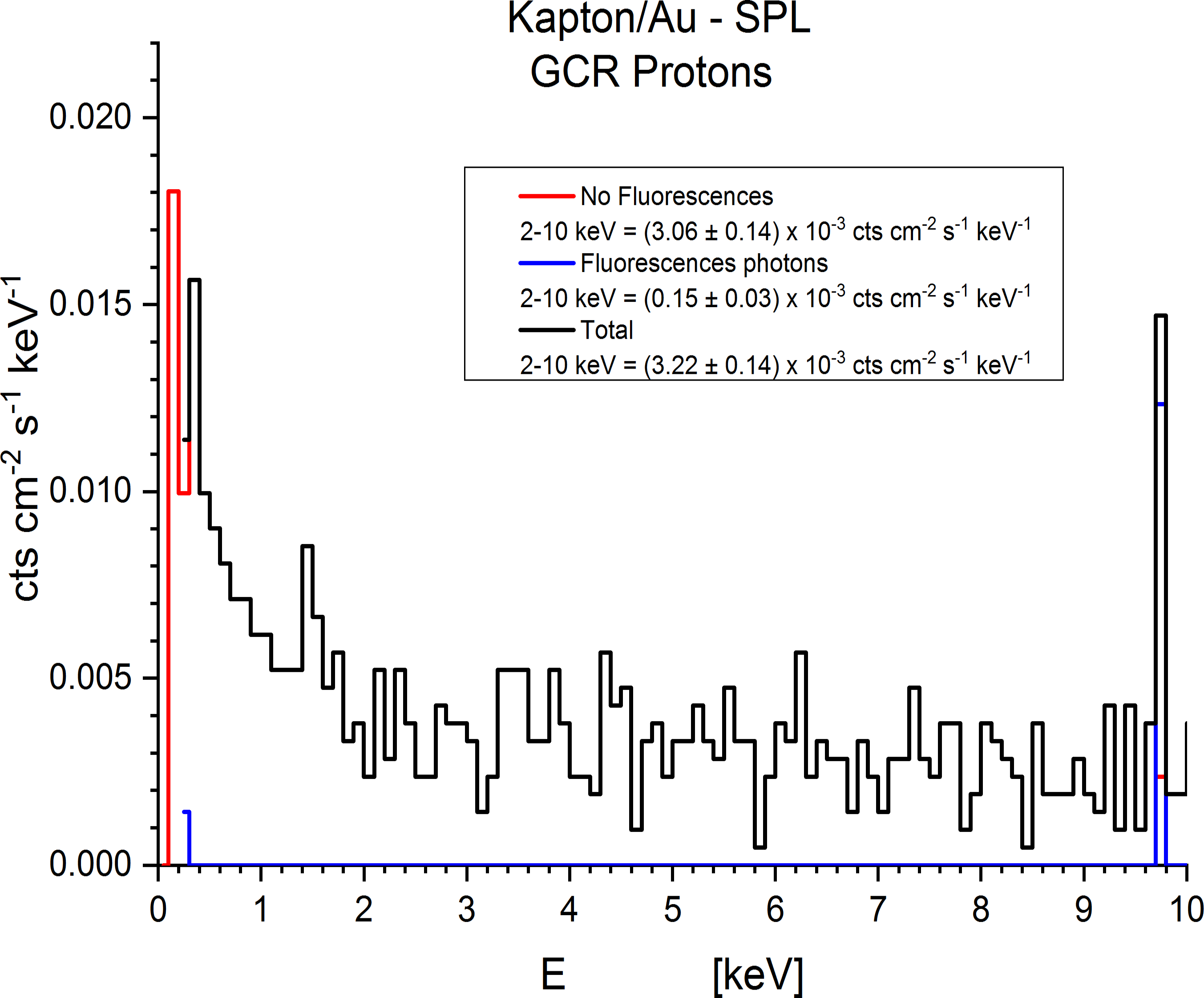}
 \caption{Unrejected particle background in the Kapton/Au shield configuration, compared with the baseline Kapton/Bi configuration (left). A more detailed analysis reveals that the difference in the integrated background value is due to the presence of the gold fluorescence line (right).}
 \label{fig:biau}
 \end{figure*}
 \begin{figure*}
 \centering
 \includegraphics[width=0.36\textwidth]{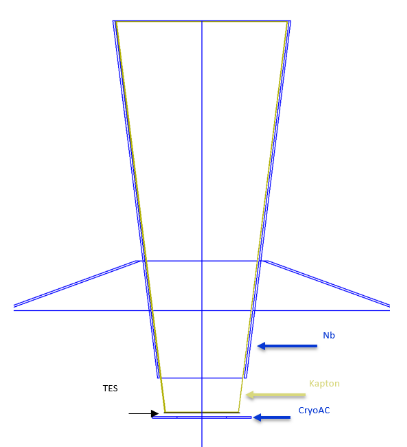} \includegraphics[width=0.6\textwidth]{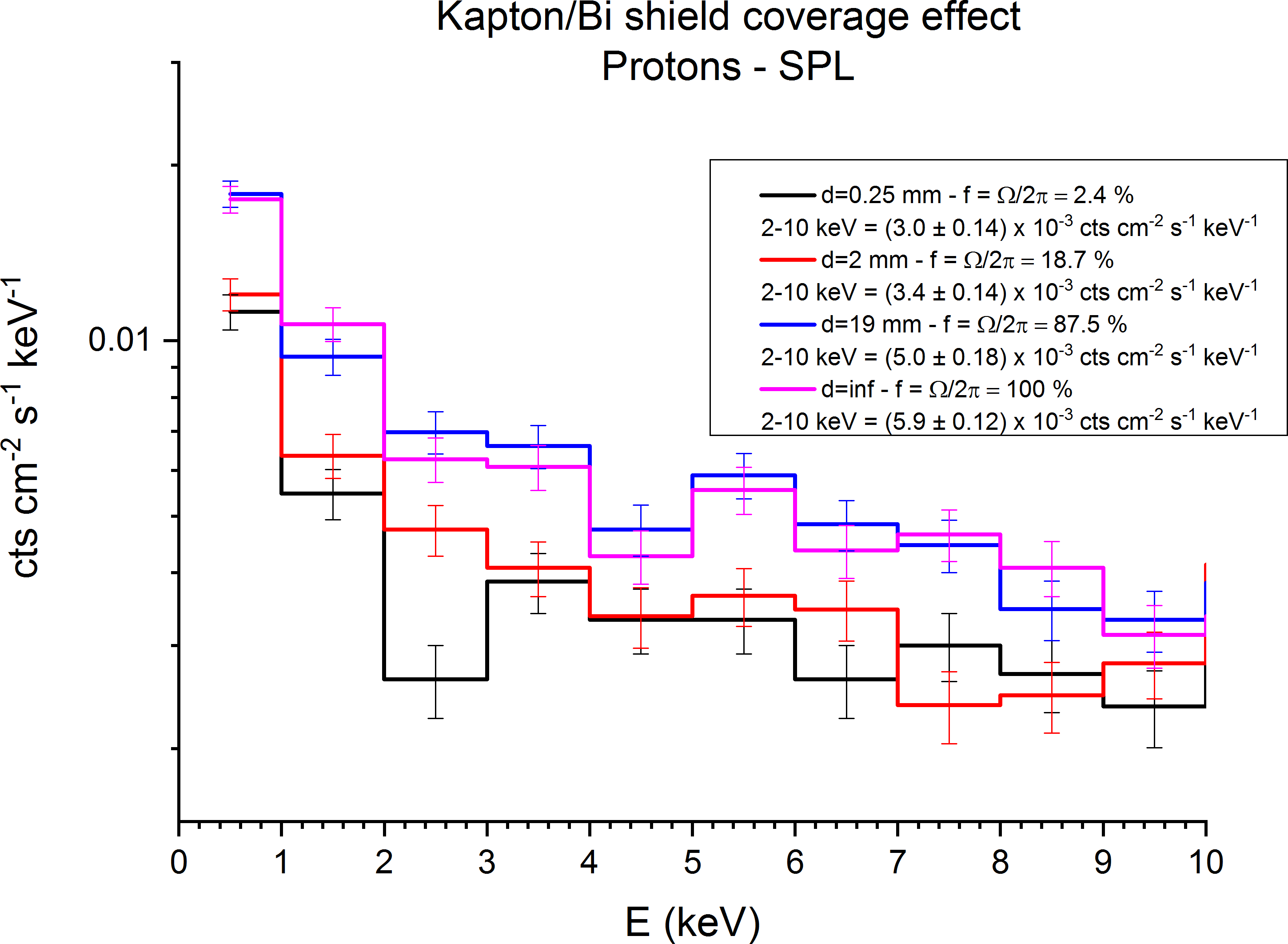}
 \caption{Drawing of the geometry in the detector proximity (left). Unrejected particle background in the Kapton/Bi configuration, for different distances of the shield from the detector surface (right).}
 \label{fig:coverage}
 \end{figure*}
 \begin{figure*}
 \centering
 \includegraphics[width=0.48\textwidth]{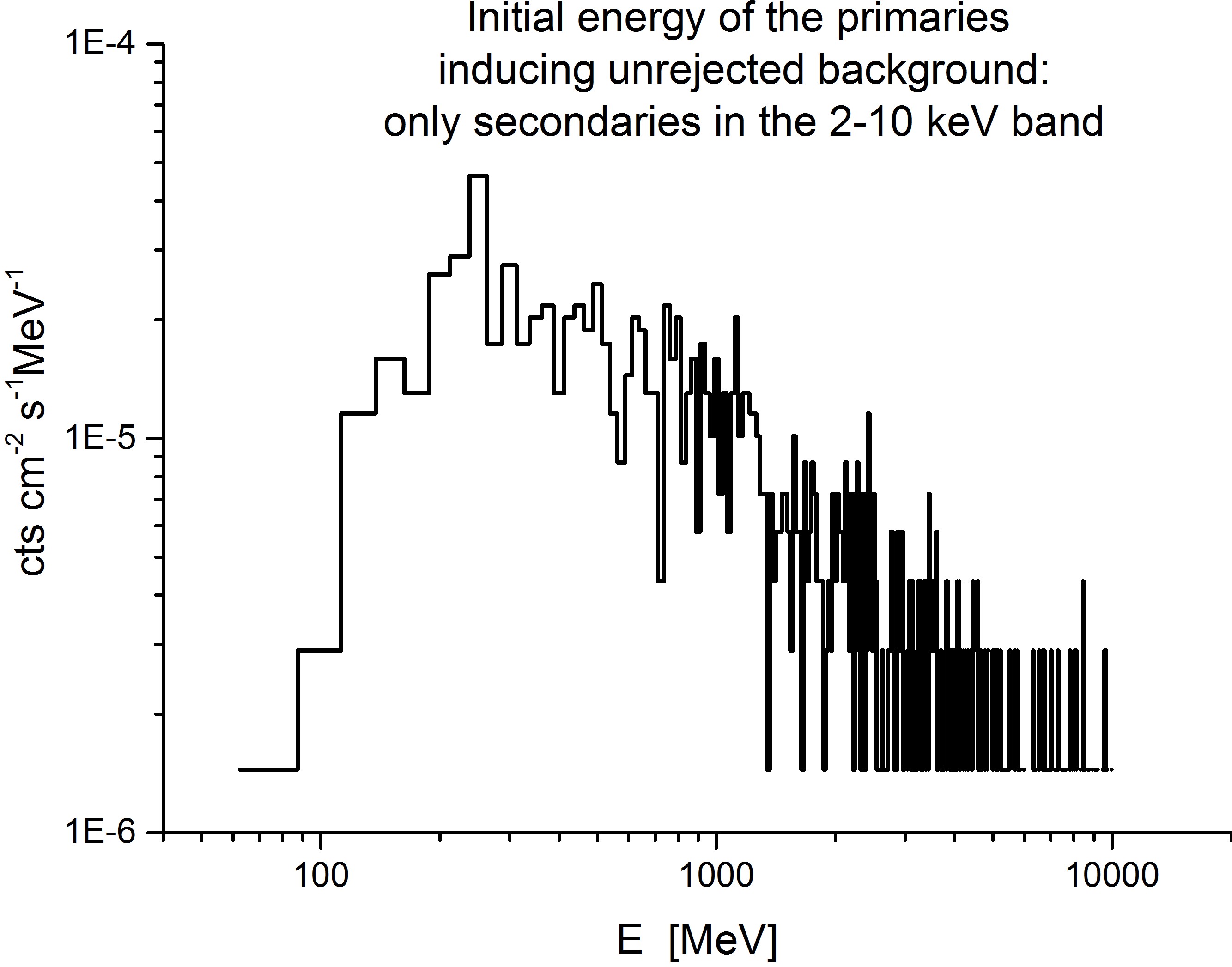} \includegraphics[width=0.48\textwidth]{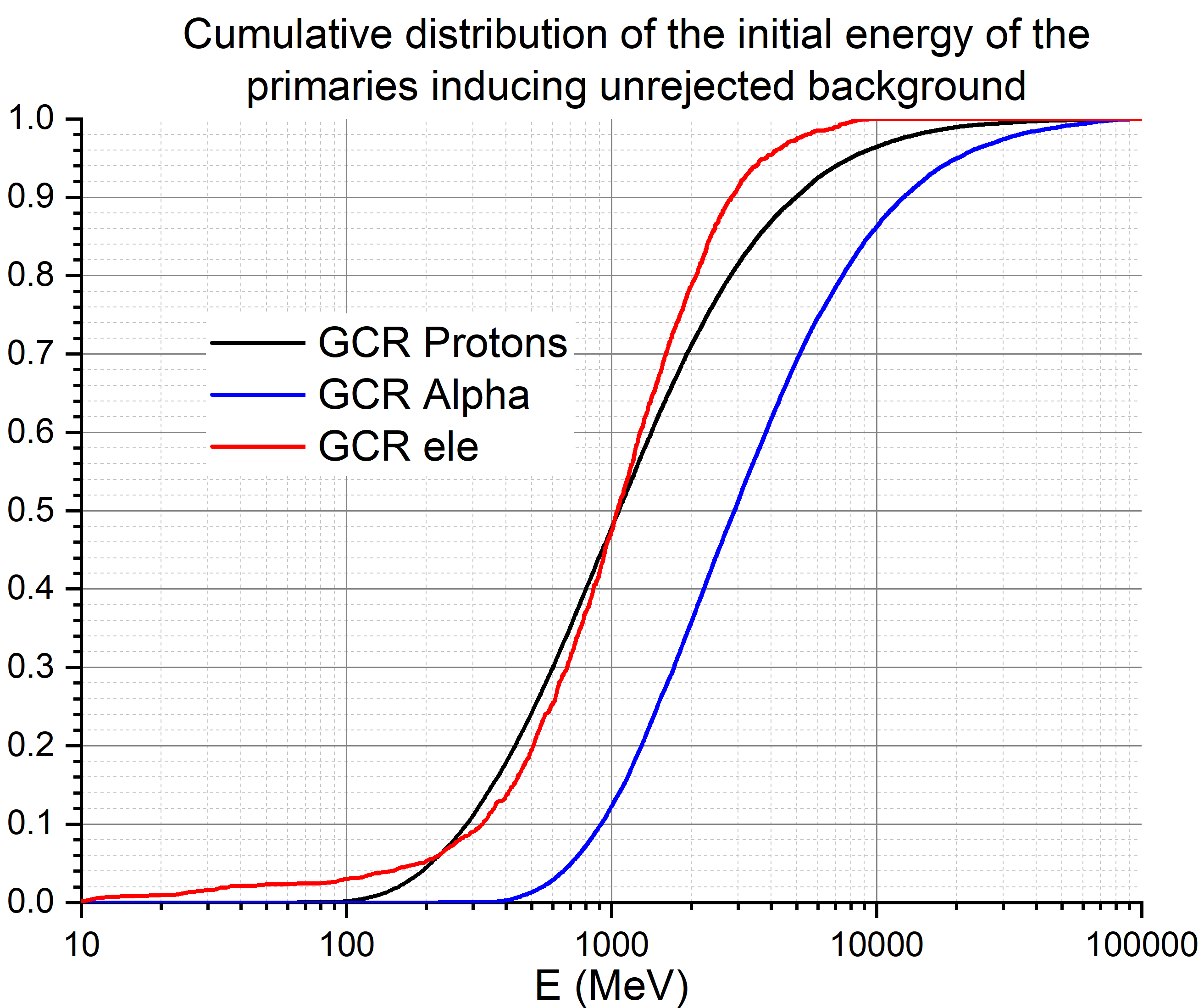}
 \caption{The initial energy of the primary particles inducing unrejected background, either directly or through the production of secondary particles (left). The cumulative plot of the curve on the left, for GCR protons, alpha particles and electrons (right).}
 \label{fig:Eini}
 \end{figure*}

With respect to the previous mass model \citep{lotti2014}, beside the revision of the structures surrounding the detector, we:
\begin{itemize}
 \item updated the TES absorbers thicknesses from 4 to 4.2 $\mu m$ for Bismuth, and from 1 to 1.7 $\mu m$ for Gold \citep{barret2018}
 \item inserted a more realistic model of the thermal filters \citep{barbera}, and the aperture cylinder sustaining them
\end{itemize}

Despite all the modifications to the mass model, the background level resulting from the GCR protons component was not affected significantly (see Figure \ref{fig:PLMMcomparison}, right).

The structures in direct sight of the detector (i.e., the Kapton/Bi passive shield and the Niobium shield) are the only ones that were not substantially modified, thus this result shows that the unrejected background value is quite robust to changes in the mass model that may happen far from the detector.

We investigated the effect of specific changes of the FPA, namely:
\begin{itemize}
 \item Modifying the thicknesses of the Kapton layer of the passive shield and of the Niobium shield inside the allowed ranges. Regarding Niobium, the unrejected background level was not influenced modifying the thickness from 500 $\mu m$ to 300 $\mu m$. Regarding Kapton, increasing the thickness to 500 $\mu m$ did not allow for any improvement of the background level, while reducing it down to 100 $\mu m$ reduced the shielding efficiency of the Kapton layer, slightly increasing the unrejected background level.
 \item Substituting the Bismuth layer with a Gold one. This was done since Bismuth is a difficult material to handle, while Gold ductility can relatively easily allow to model it into the required shape with few microns thickness. We calculated the required thickness to stop Niobium fluorescences to be 10 $\mu$m and simulated the GCR protons induced background, and results are shown in Figure \ref{fig:biau}.
 
 The unrejected background level in the Kapton/Au configuration is compatible with the one obtained in the baseline configuration if we remove the additional fluorescence line due to the presence of Gold. It should be noted that, given the narrow width of the fluorescence line and the extreme energy resolution of the X-IFU, removing the fluorescence line from the background spectrum should pose no issue, resulting in the exclusion of few bins in the response matrix of the instrument (few eV). If the fluorescence line is not removed from the spectrum, the background level increases by ~8\%, still inside 1 $\sigma$ error of present evaluation.
 
 \item Increasing the distance between the detector surface and the lower edge of the Kapton/Bi shield. In the baseline configuration the Kapton/Bi shield extends beyond the Niobium funnel (the Nb lower edge has a 19 mm distance from the detector surface) down to a distance of 0.25 mm, thus covering almost the whole solid angle the detector is exposed to (the geometry is shown in Figure \ref{fig:coverage}). This distance does not allow enough space for the detector wiring, so we tested how the residual background is affected by its increase. We tested several shield distances, and the results are reported in Figure \ref{fig:coverage}. What we found is that the background increases proportionally to the fraction of solid angle left uncovered by the shield. It should be noted that this issue can be mitigated adding a Kapton/Bi layer to the sections of the internal wall of the niobium that the detector became directly exposed to.
\end{itemize}

We also investigated the initial energy of the particles able to reach the detector we identified the energy ranges in which the different particle populations are contributing to the background (see Figure \ref{fig:Eini}), and that should be monitored by an external particle background monitor \citep{ahepam}. We can see that heavier particles require higher energies to reach the innermost part of the cryostat.

\section*{C. Particle fluxes on the detector and CryoAC rejection efficiency}\label{sec:fluxes}

 \begin{figure*}[h]
 \centering
 \includegraphics[width=0.48\textwidth]{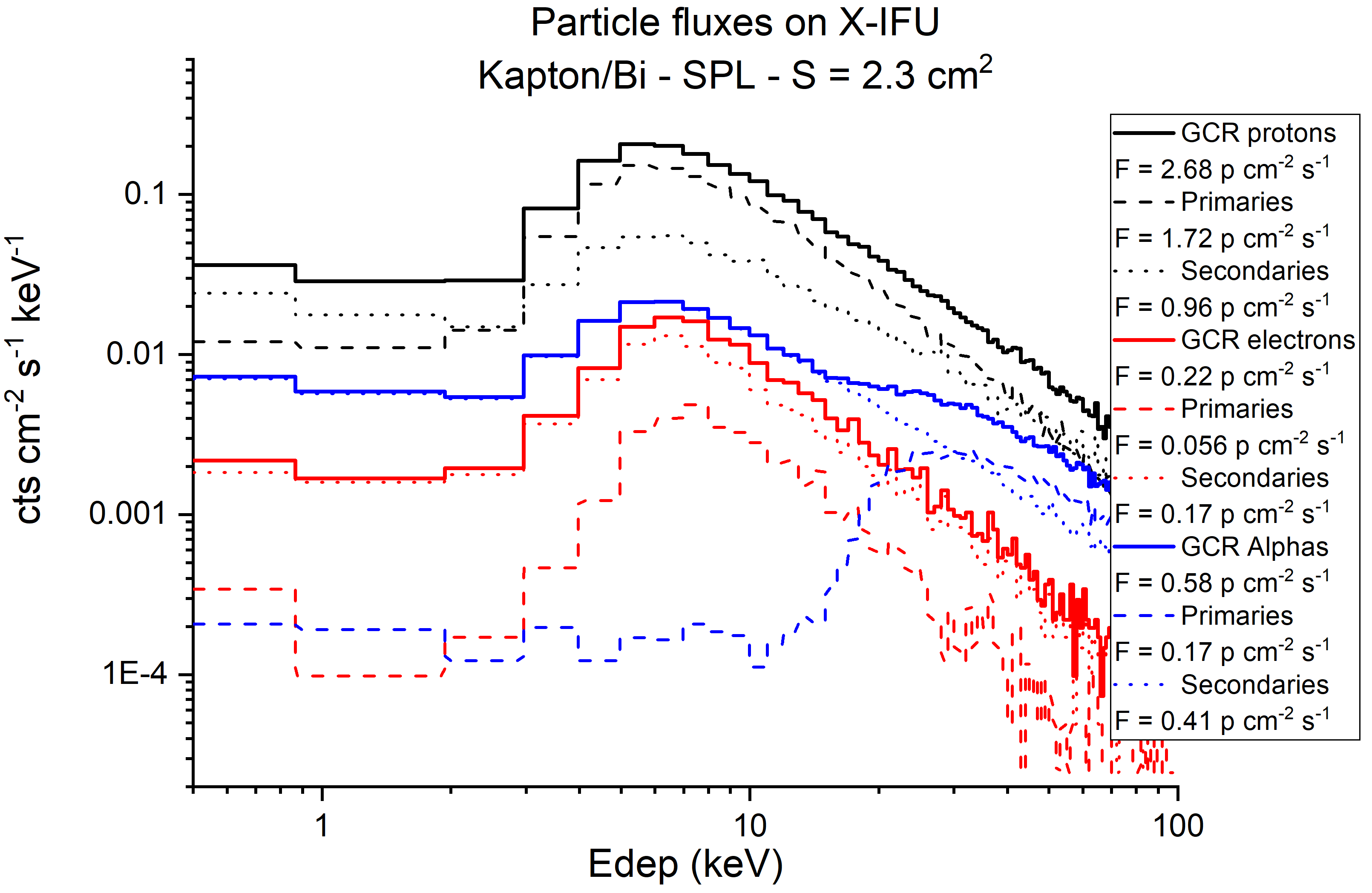} \includegraphics[width=0.48\textwidth]{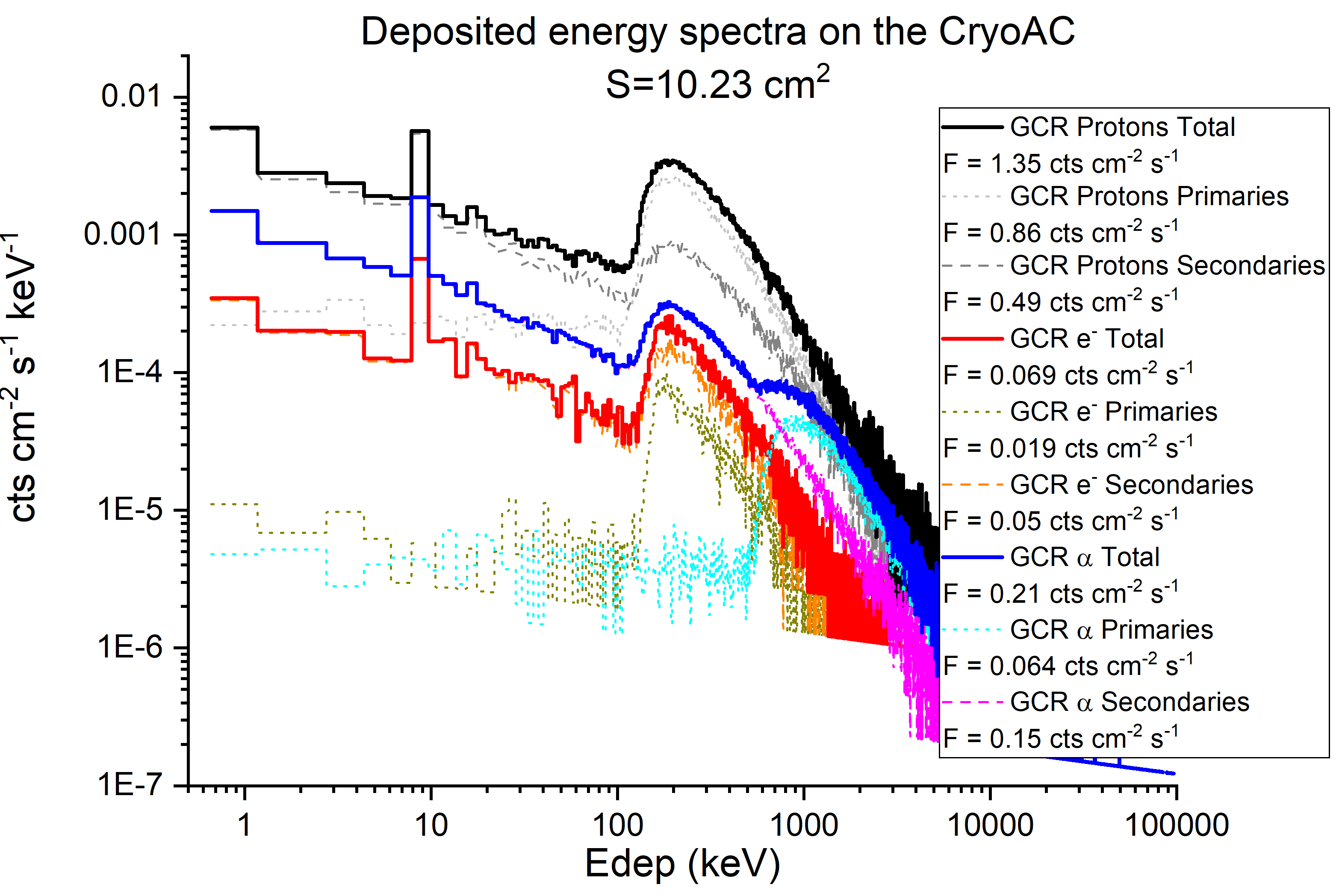} \includegraphics[width=0.48\textwidth]{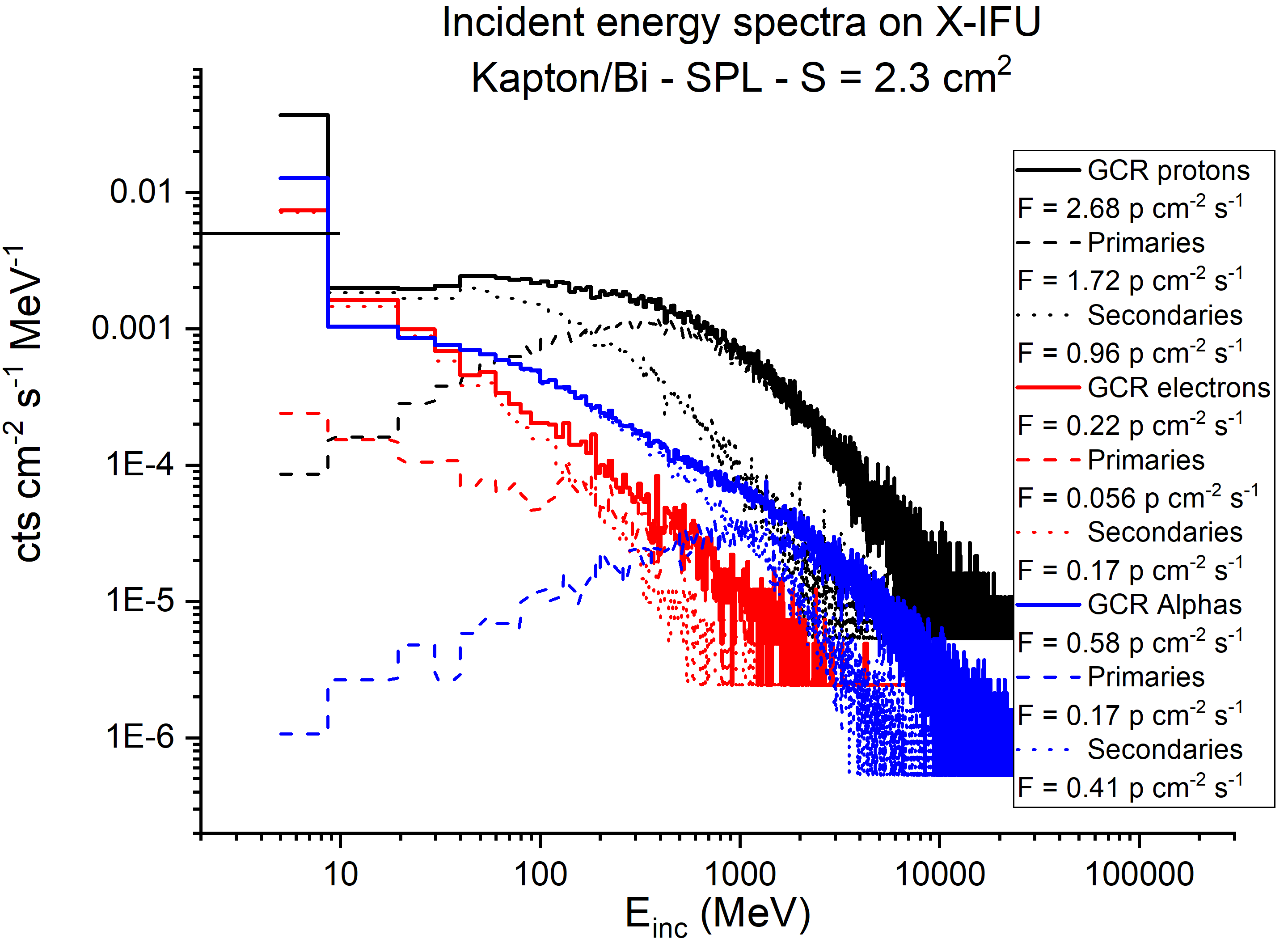} \includegraphics[width=0.48\textwidth]{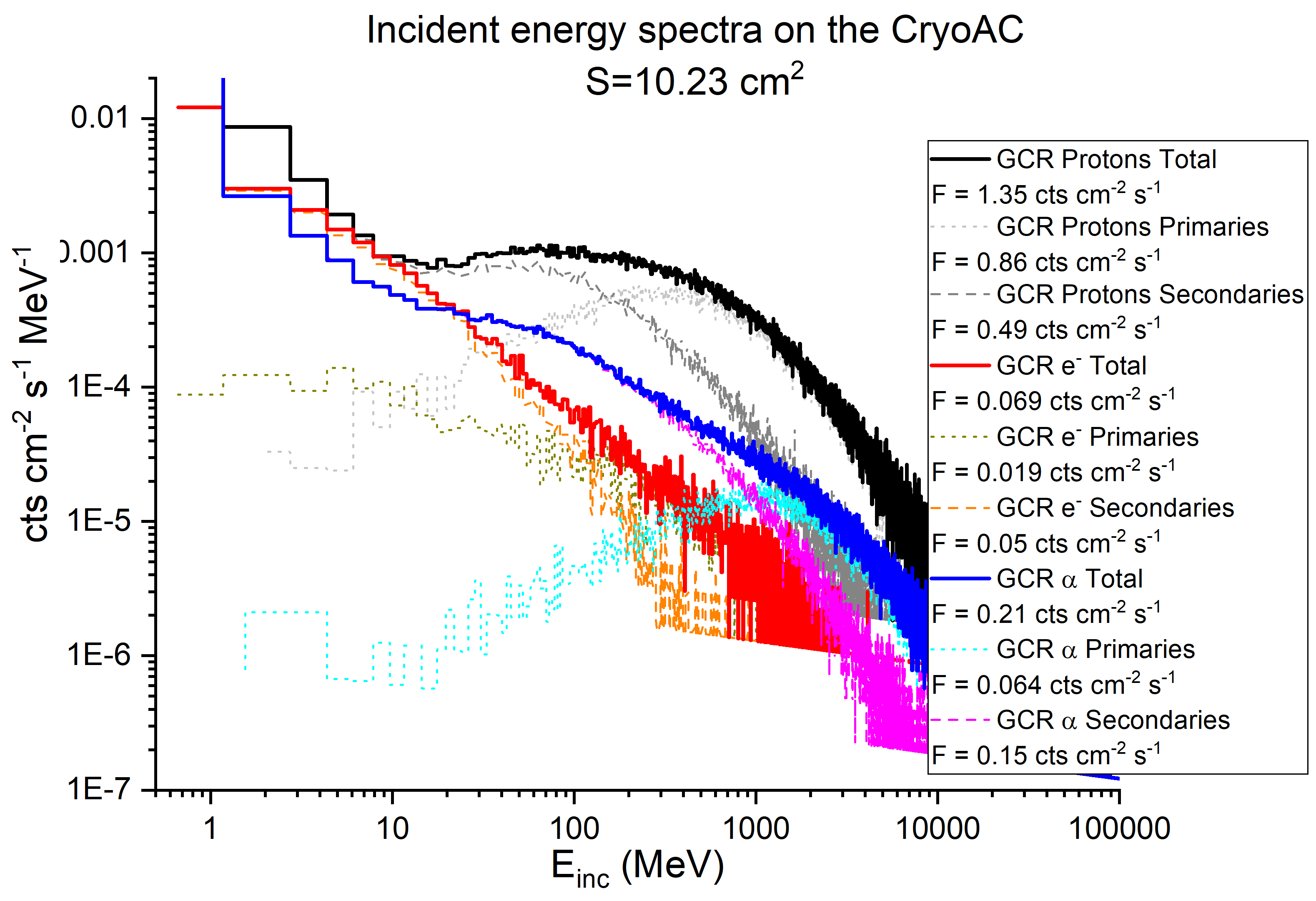}

 \caption{Top left: Deposited energy spectra and integrated fluxes in the main detector (Area=2.3 $cm^2$). Top right: Deposited energy spectra and integrated fluxes in the CryoAC (Area=10.23 $cm^2$). Note that the primary particles spectra are flat below a certain energy and do not go to zero as expected. This is due to the fact that some particles have very short paths in the CryoAC pixels. \\
 Bottom panels: incident energy spectra on the two detectors, main detector (bottom left panel, Area=2.3 $cm^2$) and CryoAC (bottom right, Area=10.23 $cm^2$); these energies are the ones possessed by the particles before interacting with the detectors. \\
 Note that the TES array detector area to be considered is the top surface, since the particle flux is to be compared to source photons flux 
 The CryoAC area is the whole surface area (top + bottom + side) since it is the flux definition. 
 }
 \label{fig:fluxes}
 \end{figure*}
 
In this section we derive the X-IFU rejection efficiency with respect to the rejectable particles analyzing the fluxes impacting on the main detector and the CryoAC. The rejection efficiency of the X-IFU system depends on the geometrical configuration of the CryoAC and main array (“geometrical rejection efficiency”, depending on distance and sizes of the two detectors) related to the classical concept of the anticoincidence solid angle covering optimization with respect to the main detector, and on the capability of the main detector array to discriminate background events on its own (i.e., autoveto), relying on the energy deposited and on the pixel pattern turned on by the impacting particles (“detector rejection efficiency”). We first derive the total rejection efficiency, that will be then broken down into its geometrical and detector efficiency. We remind that the X-IFU inefficiency is the product of the geometrical and detector inefficiencies.

To identify both the CryoAC and the main detector required rejection efficiencies in discriminating background events we first calculate the expected particle fluxes on the main detector (see Figure \ref{fig:fluxes}) for all the particle species expected in the L2 environment, separating the primary and secondary particles contribution. 

We expect on the main detector a total count rate from primary particles of 4.22 cts/s. We require this component to impact the residual background no more than 1/10 of the requirement, i.e., $5 \times 10^{-4}$ cts $cm^{-2}s^{-1}keV^{-1}$ in the 2-10 keV band, or $9.2 \times 10^{-3}$ cts/s. To reduce this flux down to the required level we require a total rejection efficiency for primary particles of $\sim99.78\%$. We remark here that the CryoAC rejection efficiency has to be calculated on the primaries component, that is entirely rejectable, and can be validated experimentally.

Analogously, we have a total count rate from secondary particles on the main detector of 2.75 cts/s. However, a fraction of this flux is induced by unrejectable particles (i.e., electrons backscattering on the detector surface releasing just a fraction of their energy, low energy particles that are completely absorbed inside the main array switching on only one pixel). The requirement on the rejection efficiency is placed on the rejectable component of the secondary particles flux, which amounts to 2.69 cts/s. As for the primaries case, we require this component to impact the residual background no more than 1/10 of the requirement, $9.2 \times 10^{-3}$ cts/s. This sets the total rejection efficiency for rejectable secondary particles to $\sim 99.66\%$.

 \begin{figure*}[h]
 \centering
 \includegraphics[width=0.48\textwidth]{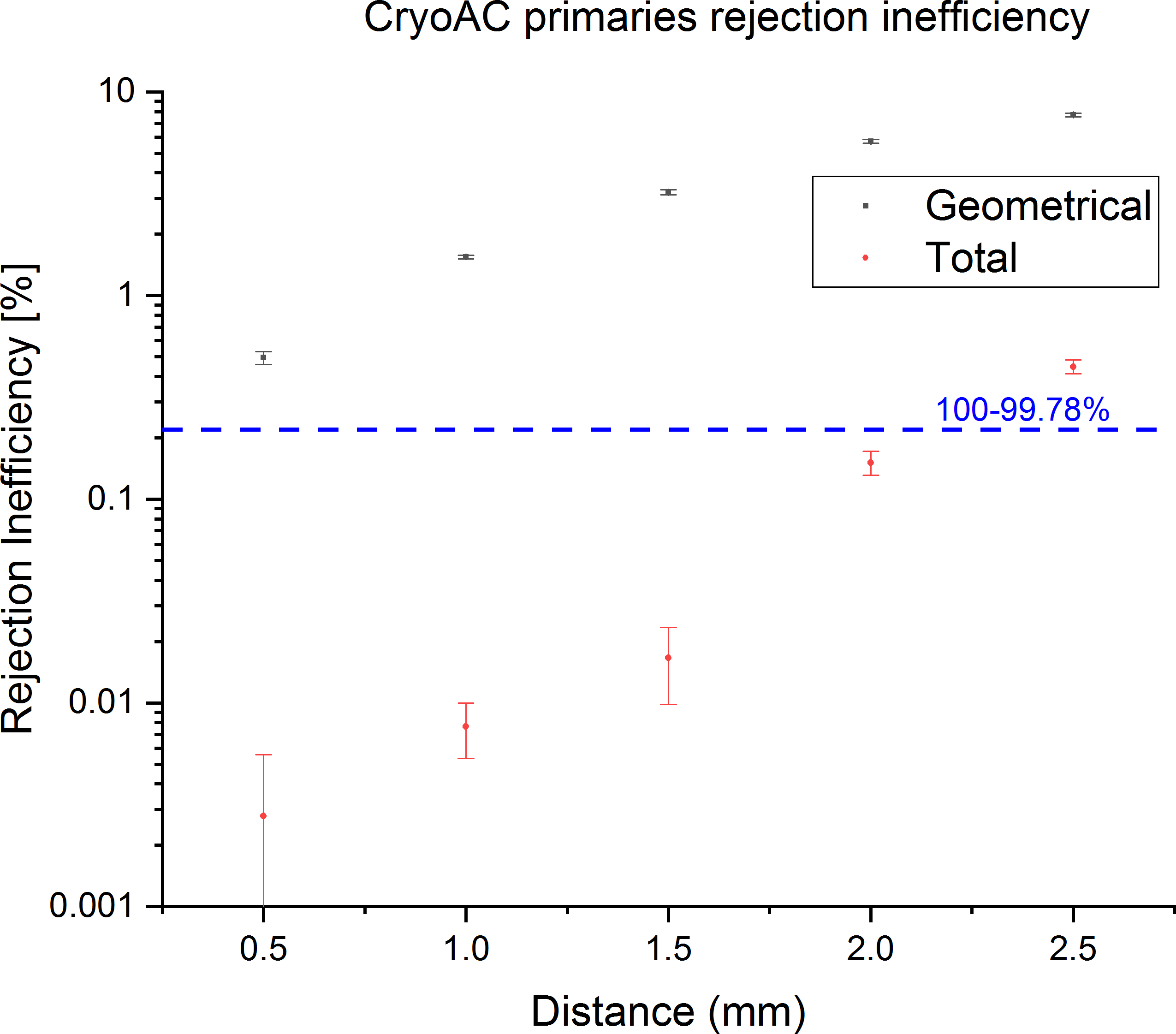} \includegraphics[width=0.48\textwidth]{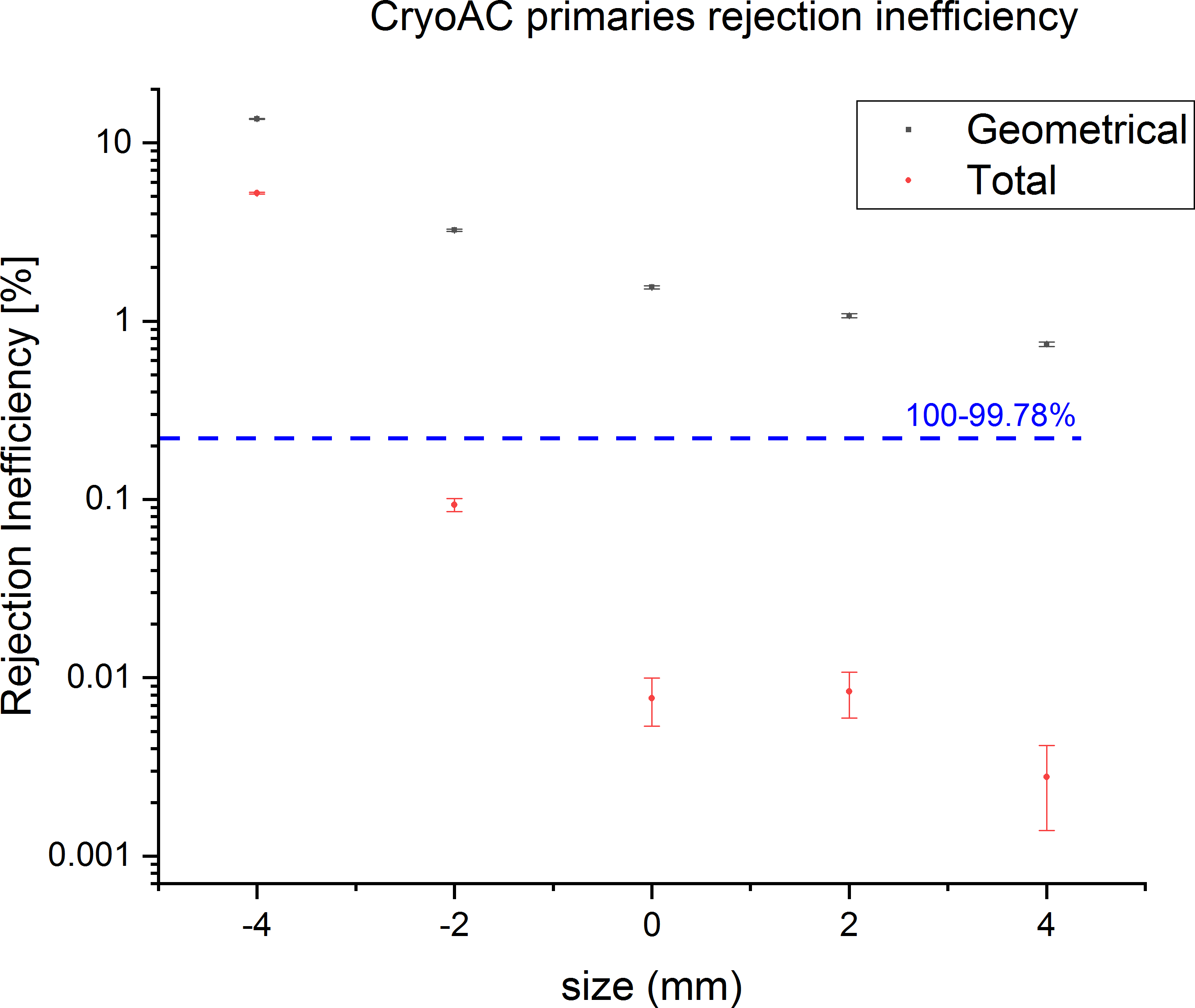}
 \caption{Geometrical and total rejection inefficiency for primary particles (defined as the unity residual of the efficiency) as function of the distance between the main detector and the CryoAC (left), and as function of the CryoAC size increment with respect to the baseline size of 11.91 mm apothem (right). We show inefficiency rather than the efficiency for clarity purpose.}
 \label{fig:rejeff}
 \end{figure*}

Now we start breaking down the X-IFU rejection efficiency evaluating the geometrical contribution. An analysis has been performed on: 
\begin{itemize}
 \item the CryoAC vs main detector related distance
 \item the CryoAC area with respect to the main detector area
\end{itemize}

we report the geometrical and total (geometrical + detector) rejection efficiencies in Figure \ref{fig:rejeff}. The results are related only to the primary component, that is entirely rejectable. Furthermore, to increase the statistics of the simulations, a reduced mass model of the instrument had to be created, and thus the secondary particles population, though not interesting for this analysis as stated above, impacting on the 2 instruments is not representative of the one experienced in the complete model. 

As it can be seen in Figure \ref{fig:rejeff}, the geometrical rejection efficiency increases with the solid angle covered by the CryoAC. A 1.5 mm distance is enough to achieve the 99.78\% total rejection efficiency goal. We adopt the 1 mm distance solution to account for some margin. Similarly, the baseline CryoAC size is sufficient to reach the goal (Figure \ref{fig:rejeff}, right), while further increase in the CryoAC size can be technically challenging and increase the instrument Dead Time, with minor benefits. The total rejection efficiency saturates beyond a given CryoAC size/distance, since we start discriminating events with very skew trajectories, that are already intrinsically discriminated by the detector rejection efficiency. 

Once the geometry of the system has been defined, we investigated the effect of the energy threshold of the CryoAC on its detection efficiency. We analyzed the residual background for the rejectable component (primary particles and secondaries). This is necessary since the primary component, despite being completely rejectable, constitutes a minor fraction of the residual background, so the secondaries contribution can’t be neglected. We found that a 20 keV threshold fulfills the requirement. Lowering further the threshold value produces no appreciable benefit, while the unrejected background value starts to rise steeply above 30 keV. 

\bibliographystyle{unsrt}  
\bibliography{test}

\begin{thebibliography}{}
\expandafter\ifx\csname natexlab\endcsname\relax\def\natexlab#1{#1}\fi
\providecommand{\url}[1]{\href{#1}{#1}}
\providecommand{\dodoi}[1]{doi:~\href{http://doi.org/#1}{\nolinkurl{#1}}}
\providecommand{\doeprint}[1]{\href{http://ascl.net/#1}{\nolinkurl{http://ascl.net/#1}}}
\providecommand{\doarXiv}[1]{\href{https://arxiv.org/abs/#1}{\nolinkurl{https://arxiv.org/abs/#1}}}

\bibitem[{Adriani {et~al.}(2013)Adriani, Barbarino, Bazilevskaya, Bellotti,
  Boezio, Bogomolov, Bongi, Bonvicini, Borisov, Bottai, Bruno, Cafagna,
  Campana, Carbone, Carlson, Casolino, Castellini, Pascale, Santis, Simone,
  Felice, Formato, Galper, Grishantseva, Karelin, Koldashov, Koldobskiy,
  Krutkov, Kvashnin, Leonov, Malakhov, Marcelli, Mayorov, Menn, Mikhailov,
  Mocchiutti, Monaco, Mori, Nikonov, Osteria, Palma, Papini, Pearce, Picozza,
  Pizzolotto, Ricci, Ricciarini, Rossetto, Sarkar, Simon, Sparvoli,
  Spillantini, Stozhkov, Vacchi, Vannuccini, Vasilyev, Voronov, Yurkin, Wu,
  Zampa, Zampa, Zverev, Potgieter, \& Vos}]{pamela}
Adriani, O., Barbarino, G.~C., Bazilevskaya, G.~A., {et~al.} 2013, The
  Astrophysical Journal, 765, 91, \dodoi{10.1088/0004-637x/765/2/91}

\bibitem[{Adriani {et~al.}(2015)Adriani, Barbarino, Bazilevskaya, Bellotti,
  Boezio, Bogomolov, Bongi, Bonvicini, Bottai, Bruno, Cafagna, Campana,
  Carlson, Casolino, Castellini, Donato, Santis, Simone, Felice, Formato,
  Galper, Karelin, Koldashov, Koldobskiy, Krutkov, Kvashnin, Leonov, Malakhov,
  Marcelli, Martucci, Mayorov, Menn, Merg{\`{e}}, Mikhailov, Mocchiutti,
  Monaco, Mori, Munini, Osteria, Palma, Panico, Papini, Pearce, Picozza, Ricci,
  Ricciarini, Sarkar, Scotti, Simon, Sparvoli, Spillantini, Stozhkov, Vacchi,
  Vannuccini, Vasilyev, Voronov, Yurkin, Zampa, Zampa, Potgieter, \&
  Vos}]{pamela-ele}
---. 2015, The Astrophysical Journal, 810, 142,
  \dodoi{10.1088/0004-637x/810/2/142}

\bibitem[{Agostinelli(2003)}]{geant1}
Agostinelli, S. e.~a. 2003, Nuclear Instruments and Methods in Physics Research
  Section A: Accelerators, Spectrometers, Detectors and Associated Equipment,
  Volume 506, \dodoi{10.1016/S0168-9002(03)01368-8}

\bibitem[{Allison(2006)}]{geant2}
Allison, J. e.~a. 2006, IEEE, 53, \dodoi{10.1109/TNS.2006.869826}

\bibitem[{Allison(2016)}]{geant3}
---. 2016, Nuclear Instruments and Methods in Physics Research Section A:
  Accelerators, Spectrometers, Detectors and Associated Equipment, 835,
  \dodoi{10.1016/j.nima.2016.06.125}

\bibitem[{{Anders} \& {Grevesse}(1989)}]{andersgrevesse}
{Anders}, E., \& {Grevesse}, N. 1989, Geochimica et Cosmochimica Acta, 53, 197,
  \dodoi{10.1016/0016-7037(89)90286-X}

\bibitem[{Angelopoulos(2011)}]{artemis}
Angelopoulos, V. 2011, Space Science Reviews, 165, 3,
  \dodoi{10.1007/s11214-010-9687-2}

\bibitem[{{Arnaud}(1996)}]{xspec}
{Arnaud}, K.~A. 1996, Astronomical Society of the Pacific Conference Series,
  Vol. 101, {XSPEC: The First Ten Years}, ed. G.~H. {Jacoby} \& J.~{Barnes}, 17

\bibitem[{Barbera(2016)}]{barbera}
Barbera, M., A. A. B. E. e.~a. 2016, Journal of Low Temperature Physics, 184,
  \dodoi{10.1007/s10909-016-1501-4}

\bibitem[{Barret \& Cucchetti(2018)}]{respmatrix}
Barret, D., \& Cucchetti, E. 2018, X-IFU response matrices, online database.
\newblock
  \url{http://x-ifu-resources.irap.omp.eu/PUBLIC/RESPONSES/CC_CONFIGURATION/}

\bibitem[{Barret {et~al.}(2018)Barret, Trong, den Herder, Piro, Cappi,
  Houvelin, Kelley, Mas-Hesse, Mitsuda, Paltani, Rauw, Rozanska, Wilms,
  Bandler, Barbera, Barcons, Bozzo, Ceballos, Charles, Costantini,
  Decourchelle, den Hartog, Duband, Duval, Fiore, Gatti, Goldwurm, Jackson,
  Jonker, Kilbourne, Macculi, Mendez, Molendi, Orleanski, Pajot, Pointecouteau,
  Porter, Pratt, Prêle, Ravera, Sato, Schaye, Shinozaki, Thibert, Valenziano,
  Valette, Vink, Webb, Wise, Yamasaki, Douchin, Mesnager, Pontet, Pradines,
  Branduardi-Raymont, Bulbul, Dadina, Ettori, Finoguenov, Fukazawa, Janiuk,
  Kaastra, Mazzotta, Miller, Miniutti, Naze, Nicastro, Scioritino, Simonescu,
  Torrejon, Frezouls, Geoffray, Peille, Aicardi, André, Daniel, Clénet,
  Etcheverry, Gloaguen, Hervet, Jolly, Ledot, Paillet, Schmisser, Vella,
  Damery, Boyce, Dipirro, Lotti, Schwander, Smith, Leeuwen, van Weers, Clerc,
  Cobo, Dauser, Kirsch, Cucchetti, Eckart, Ferrando, \& Natalucci}]{barret2018}
Barret, D., Trong, T.~L., den Herder, J.-W., {et~al.} 2018, in Space Telescopes
  and Instrumentation 2018: Ultraviolet to Gamma Ray, ed. J.-W.~A. den Herder,
  S.~Nikzad, \& K.~Nakazawa, Vol. 10699, International Society for Optics and
  Photonics (SPIE), 324 -- 338, \dodoi{10.1117/12.2312409}

\bibitem[{Boschini {et~al.}(2019)Boschini, Torre, Gervasi, Vacca, \&
  Rancoita}]{helmod}
Boschini, M., Torre, S.~D., Gervasi, M., Vacca, G.~L., \& Rancoita, P. 2019,
  Advances in Space Research, 64, 2459 , \dodoi{10.1016/j.asr.2019.04.007}

\bibitem[{Boschini {et~al.}(2018)Boschini, Torre, Gervasi, Grandi,
  Jóhannesson, Vacca, Masi, Moskalenko, Pensotti, Porter, Quadrani, Rancoita,
  Rozza, \& Tacconi}]{helmodalpha}
Boschini, M.~J., Torre, S.~D., Gervasi, M., {et~al.} 2018, Astrophys. J., 858,
  61, \dodoi{10.3847/1538-4357/aabc54}

\bibitem[{{Boschini} {et~al.}(2018){Boschini}, {Della Torre}, {Gervasi},
  {Grandi}, J{\'o}hannesson, {La Vacca}, {Masi}, {Moskalenko}, {Pensotti},
  {Porter}, {Quadrani}, {Rancoita}, {Rozza}, \& {Tacconi}}]{helmodele}
{Boschini}, M.~J., {Della Torre}, S., {Gervasi}, M., {et~al.} 2018, Astrophys.
  J., 854, 94, \dodoi{10.3847/1538-4357/aaa75e}

\bibitem[{Boschini {et~al.}(2020)Boschini, Torre, Gervasi, Grandi,
  Jóhannesson, Vacca, Masi, Moskalenko, Pensotti, Porter, Quadrani, Rancoita,
  Rozza, \& Tacconi}]{helmodalpha2}
Boschini, M.~J., Torre, S.~D., Gervasi, M., {et~al.} 2020, Astrophys. J., 889,
  167, \dodoi{10.3847/1538-4357/ab64f1}

\bibitem[{Cucchetti {et~al.}(2018)Cucchetti, Pointecouteau, Peille, Clerc,
  Rasia, Biffi, Borgani, Dolag, Wilms, Pajot, \& Barret}]{cucchetti}
Cucchetti, E., Pointecouteau, E., Peille, P., {et~al.} 2018, in Space
  Telescopes and Instrumentation 2018: Ultraviolet to Gamma Ray, ed. J.-W.~A.
  den Herder, S.~Nikzad, \& K.~Nakazawa, Vol. 10699, International Society for
  Optics and Photonics (SPIE), 1119 -- 1129, \dodoi{10.1117/12.2311957}

\bibitem[{{Cucchetti, E.} {et~al.}(2019){Cucchetti, E.}, {Clerc, N.},
  {Pointecouteau, E.}, {Peille, P.}, \& {Pajot, F.}}]{cucchetti2}
{Cucchetti, E.}, {Clerc, N.}, {Pointecouteau, E.}, {Peille, P.}, \& {Pajot, F.}
  2019, A\&A, 629, A144, \dodoi{10.1051/0004-6361/201935677}

\bibitem[{Eckert(2012)}]{eckert2012}
Eckert, D. e.~a. 2012, Astronomy \& Astrophysics, 541,
  \dodoi{10.1051/0004-6361/201118281}

\bibitem[{Ettori {et~al.}(2013)Ettori, Pratt, de~Plaa, Eckert, Nevalainen,
  Battistelli, Borgani, Croston, Finoguenov, Kaastra, Gaspari, Gastaldello,
  Gitti, Molendi, Pointecouteau, Ponman, Reiprich, Roncarelli, Rossetti,
  Sanders, Sun, Trinchieri, Vazza, Arnaud, Böringher, Brighenti, Dahle,
  Grandi, Mohr, Moretti, \& Schindler}]{ettori2013hot}
Ettori, S., Pratt, G.~W., de~Plaa, J., {et~al.} 2013, The Hot and Energetic
  Universe: The astrophysics of galaxy groups and clusters.
\newblock \doarXiv{1306.2322}

\bibitem[{{Fioretti} {et~al.}(2012){Fioretti}, {Bulgarelli}, {Malaguti},
  {Bianchin}, {Trifoglio}, \& {Gianotti}}]{normalization3}
{Fioretti}, V., {Bulgarelli}, A., {Malaguti}, G., {et~al.} 2012, in Society of
  Photo-Optical Instrumentation Engineers (SPIE) Conference Series, Vol. 8453,
  High Energy, Optical, and Infrared Detectors for Astronomy V, ed. A.~D.
  {Holland} \& J.~W. {Beletic}, 845331, \dodoi{10.1117/12.926248}

\bibitem[{Foster {et~al.}(2012)Foster, Ji, Smith, \& Brickhouse}]{apec}
Foster, A.~R., Ji, L., Smith, R.~K., \& Brickhouse, N.~S. 2012, The
  Astrophysical Journal, 756, 128, \dodoi{10.1088/0004-637x/756/2/128}

\bibitem[{Ghirardini(2019)}]{ghirardini2019}
Ghirardini, V. e.~a. 2019, Astronomy and Astrophysics, 627,
  \dodoi{10.1051/0004-6361/201834875}

\bibitem[{{Ghizzardi} {et~al.}(2021){Ghizzardi}, {Molendi}, {Van der Burg}, R.,
  {Bartalucci}, {Gastaldello}, {Rossetti}, {Biffi}, {Borgani}, {Eckert},
  {Ettori}, {Gaspari}, {Ghirardini}, \& {Rasia}}]{ghizzardi2021}
{Ghizzardi}, S., {Molendi}, S., {Van der Burg}, {et~al.} 2021, A\&A.
\newblock \doarXiv{2007.01084}

\bibitem[{Grimani {et~al.}(2009)Grimani, Fabi, Finetti, \&
  Tombolato}]{Grimani2009}
Grimani, C., Fabi, M., Finetti, N., \& Tombolato, D. 2009, Classical and
  Quantum Gravity, 26, 215004, \dodoi{10.1088/0264-9381/26/21/215004}

\bibitem[{{K{\"u}hl} {et~al.}(2016){K{\"u}hl}, {G{\'o}mez-Herrero}, \&
  {Heber}}]{SOHO}
{K{\"u}hl}, P., {G{\'o}mez-Herrero}, R., \& {Heber}, B. 2016, Solar Physics,
  291, 965, \dodoi{10.1007/s11207-016-0879-0}

\bibitem[{Kuznetsov {et~al.}(2017)Kuznetsov, Popova, \& Panasyuk}]{alpha}
Kuznetsov, N.~V., Popova, H., \& Panasyuk, M.~I. 2017, Journal of Geophysical
  Research: Space Physics, 122, 1463, \dodoi{10.1002/2016JA022920}

\bibitem[{Laurenza(2019)}]{laurenza2019}
Laurenza, M. e.~a. 2019, The Astrophysical Journal, 873,
  \dodoi{10.3847/1538-4357/ab0410}

\bibitem[{{Leccardi} \& {Molendi}(2008{\natexlab{a}})}]{leccardimolendi2008}
{Leccardi}, A., \& {Molendi}, S. 2008{\natexlab{a}}, Astronomy \& Astrophysics,
  487, 461, \dodoi{10.1051/0004-6361:200810113}

\bibitem[{{Leccardi} \& {Molendi}(2008{\natexlab{b}})}]{leccardimolendi2008-2}
---. 2008{\natexlab{b}}, Astronomy \& Astrophysics, 486, 359,
  \dodoi{10.1051/0004-6361:200809538}

\bibitem[{Lotti {et~al.}(2018)Lotti, Macculi, D'Andrea, Fioretti, Dondero,
  Mantero, Minervini, Argan, \& Piro}]{lotti2018spie}
Lotti, S., Macculi, C., D'Andrea, M., {et~al.} 2018, in Space Telescopes and
  Instrumentation 2018: Ultraviolet to Gamma Ray, ed. J.-W.~A. den Herder,
  S.~Nikzad, \& K.~Nakazawa, Vol. 10699, International Society for Optics and
  Photonics (SPIE), 397 -- 405, \dodoi{10.1117/12.2313236}

\bibitem[{Lotti(2014)}]{lotti2014}
Lotti, S. e.~a. 2014, A\&A, 569, \dodoi{10.1051/0004-6361/201323307}

\bibitem[{Lotti(2017)}]{lotti2017}
---. 2017, Experimental Astronomy, \dodoi{10.1007/s10686-017-9538-1}

\bibitem[{Lotti(2018)}]{lotti2018}
---. 2018, Experimental Astronomy, 45, \dodoi{10.1007/s10686-018-9599-9}

\bibitem[{Meidinger {et~al.}(2018)Meidinger, Nandra, \& Plattner}]{wfi}
Meidinger, N., Nandra, K., \& Plattner, M. 2018, in Space Telescopes and
  Instrumentation 2018: Ultraviolet to Gamma Ray, ed. J.-W.~A. den Herder,
  S.~Nikzad, \& K.~Nakazawa, Vol. 10699, International Society for Optics and
  Photonics (SPIE), 312 -- 323, \dodoi{10.1117/12.2310141}

\bibitem[{Molendi(2016)}]{molendi2016}
Molendi, S. e.~a. 2016, Astronomy \& Astrophysics, 586,
  \dodoi{10.1051/0004-6361/201527356}

\bibitem[{Molendi(2018)}]{ahepam}
---. 2018.
\newblock \url{https://1drv.ms/b/s!AhxWjpBdUfEJgf4IqS0lAFVCR2wIWw}

\bibitem[{Nandra {et~al.}(2013)Nandra, Barret, Barcons, Fabian, den Herder,
  Piro, Watson, Adami, Aird, Afonso, Alexander, Argiroffi, Amati, Arnaud,
  Atteia, Audard, Badenes, Ballet, Ballo, Bamba, Bhardwaj, Battistelli, Becker,
  Becker, Behar, Bianchi, Biffi, Bîrzan, Bocchino, Bogdanov, Boirin, Boller,
  Borgani, Borm, Bouché, Bourdin, Bower, Braito, Branchini,
  Branduardi-Raymont, Bregman, Brenneman, Brightman, Brüggen, Buchner, Bulbul,
  Brusa, Bursa, Caccianiga, Cackett, Campana, Cappelluti, Cappi, Carrera,
  Ceballos, Christensen, Chu, Churazov, Clerc, Corbel, Corral, Comastri,
  Costantini, Croston, Dadina, D'Ai, Decourchelle, Ceca, Dennerl, Dolag, Done,
  Dovciak, Drake, Eckert, Edge, Ettori, Ezoe, Feigelson, Fender, Feruglio,
  Finoguenov, Fiore, Galeazzi, Gallagher, Gandhi, Gaspari, Gastaldello,
  Georgakakis, Georgantopoulos, Gilfanov, Gitti, Gladstone, Goosmann, Gosset,
  Grosso, Guedel, Guerrero, Haberl, Hardcastle, Heinz, Herrero, Hervé,
  Holmstrom, Iwasawa, Jonker, Kaastra, Kara, Karas, Kastner, King, Kosenko,
  Koutroumpa, Kraft, Kreykenbohm, Lallement, Lanzuisi, Lee, Lemoine-Goumard,
  Lobban, Lodato, Lovisari, Lotti, McCharthy, McNamara, Maggio, Maiolino,
  Marco, de~Martino, Mateos, Matt, Maughan, Mazzotta, Mendez, Merloni, Micela,
  Miceli, Mignani, Miller, Miniutti, Molendi, Montez, Moretti, Motch, Nazé,
  Nevalainen, Nicastro, Nulsen, Ohashi, O'Brien, Osborne, Oskinova, Pacaud,
  Paerels, Page, Papadakis, Pareschi, Petre, Petrucci, Piconcelli, Pillitteri,
  Pinto, de~Plaa, Pointecouteau, Ponman, Ponti, Porquet, Pounds, Pratt,
  Predehl, Proga, Psaltis, Rafferty, Ramos-Ceja, Ranalli, Rasia, Rau, Rauw,
  Rea, Read, Reeves, Reiprich, Renaud, Reynolds, Risaliti, Rodriguez, Hidalgo,
  Roncarelli, Rosario, Rossetti, Rozanska, Rovilos, Salvaterra, Salvato, Salvo,
  Sanders, Sanz-Forcada, Schawinski, Schaye, Schwope, Sciortino, Severgnini,
  Shankar, Sijacki, Sim, Schmid, Smith, Steiner, Stelzer, Stewart, Strohmayer,
  Strüder, Sun, Takei, Tatischeff, Tiengo, Tombesi, Trinchieri, Tsuru,
  Ud-Doula, Ursino, Valencic, Vanzella, Vaughan, Vignali, Vink, Vito,
  Volonteri, Wang, Webb, Willingale, Wilms, Wise, Worrall, Young, Zampieri,
  Zand, Zane, Zezas, Zhang, \& Zhuravleva}]{whitepaper}
Nandra, K., Barret, D., Barcons, X., {et~al.} 2013, The Hot and Energetic
  Universe: A White Paper presenting the science theme motivating the Athena+
  mission.
\newblock \doarXiv{1306.2307}

\bibitem[{Ozaki \& Fioretti(2018)}]{fioretti2018}
Ozaki, M., \& Fioretti, V. 2018, Hitomi-related Geant4 activities.
\newblock
  \url{https://indico.esa.int/event/249/contributions/4195/attachments/3262/4235/Hitomi-related_Geant4_activities.pdf}

\bibitem[{Pointecouteau {et~al.}(2013)Pointecouteau, Reiprich, Adami, Arnaud,
  Biffi, Borgani, Borm, Bourdin, Brueggen, Bulbul, Clerc, Croston, Dolag,
  Ettori, Finoguenov, Kaastra, Lovisari, Maughan, Mazzotta, Pacaud, de~Plaa,
  Pratt, Ramos-Ceja, Rasia, Sanders, Zhang, Allen, Boehringer, Brunetti, Elbaz,
  Fassbender, Hoekstra, Hildebrandt, Lamer, Marrone, Mohr, Molendi, Nevalainen,
  Ohashi, Ota, Pierre, Romer, Schindler, Schrabback, Schwope, Smith, Springel,
  \& von~der Linden}]{pointecouteau2013hot}
Pointecouteau, E., Reiprich, T.~H., Adami, C., {et~al.} 2013, The Hot and
  Energetic Universe: The evolution of galaxy groups and clusters.
\newblock \doarXiv{1306.2319}

\bibitem[{{Roncarelli, M.} {et~al.}(2018){Roncarelli, M.}, {Gaspari, M.},
  {Ettori, S.}, {Biffi, V.}, {Brighenti, F.}, {Bulbul, E.}, {Clerc, N.},
  {Cucchetti, E.}, {Pointecouteau, E.}, \& {Rasia, E.}}]{Roncarelli}
{Roncarelli, M.}, {Gaspari, M.}, {Ettori, S.}, {et~al.} 2018, A\&A, 618, A39,
  \dodoi{10.1051/0004-6361/201833371}

\bibitem[{Usoskin(2001)}]{oulu}
Usoskin, I. 2001, Cosmic Ray Station of the University of Oulu, On-Line
  Database of Cosmic Ray Intensities.
\newblock \url{https://ui.adsabs.harvard.edu/abs/2001ICRC....9.3842U/abstract}

\bibitem[{Usoskin {et~al.}(2005)Usoskin, Alanko-Huotari, Kovaltsov, \&
  Mursula}]{usoskin2005}
Usoskin, I.~G., Alanko-Huotari, K., Kovaltsov, G.~A., \& Mursula, K. 2005,
  Journal of Geophysical Research: Space Physics, 110,
  \dodoi{10.1029/2005JA011250}

\bibitem[{{Verner} {et~al.}(1996){Verner}, {Ferland}, {Korista}, \&
  {Yakovlev}}]{phabs}
{Verner}, D.~A., {Ferland}, G.~J., {Korista}, K.~T., \& {Yakovlev}, D.~G. 1996,
  The Astrophysical Journal, 465, 487, \dodoi{10.1086/177435}

\end{thebibliography}

\end{document}